\documentclass[prx,twocolumn,bibnotes,showpacs,amsbsy,floatfix,amsbsy,floatfix,superscriptaddress]{revtex4-1}
\usepackage{epsfig,color}
\usepackage[cp1250]{inputenc}
\usepackage{amsmath}
\usepackage{amssymb}
\usepackage{amsfonts}
\usepackage{color}
\usepackage{wasysym}
\usepackage{latexsym} 
\usepackage[dvipsnames]{xcolor}
\usepackage{float}
\usepackage{graphicx,color}

\begin{document}
\title{Spin-resolved nonlocal transport in proximitized Rashba nanowires}
\author{Pawe\l{} Szumniak}
\affiliation{AGH University of Krakow, Faculty of Physics and Applied Computer Science, Al. Mickiewicza 30, 30-059 Krak\'ow, Poland}
\author{Daniel Loss}
\affiliation{Department of Physics, University of Basel, Klingelbergstrasse 82, 4056 Basel, Switzerland}
\author{Jelena Klinovaja}
\affiliation{Department of Physics, University of Basel, Klingelbergstrasse 82, 4056 Basel, Switzerland}

\begin{abstract}
Non-equilibrium transport in hybrid semiconductor-superconductor nanowires is crucial for many quantum phenomena such as generating entangled states via cross Andreev reflection (CAR) processes, detecting topological superconductivity, reading out Andreev spin qubits, coupling spin qubits over long distances and so on. Here, we investigate numerically transport properties of a proximitized Rashba nanowire that hosts spin-polarized low-energy quasiparticle states. We show that the spin polarization in such one-dimensional Andreev bands, extended over the entire nanowire length, can be detected in nonlocal transport measurements with tunnel-coupled side leads that are spin polarized. Remarkably, we find an exact  correspondence between the sign of the nonlocal conductance and the spin density of  the superconducting quasiparticles at the side lead position. We demonstrate that this feature is robust to moderate static disorder. As an example, we show that such a method can be used to detect spin inversion of the bands, accompanying the topological phase transition (TPT) for realistic system parameters. Furthermore, we show that such effects can be used to switch between CAR and elastic cotunneling (ECT) processes by tuning the strength of either the electric or the magnetic field. These findings hold significant practical implications for state-of-the-art transport experiments in such hybrid systems. 
\end{abstract}
\maketitle

\section{Introduction} 
Superconductor-semiconductor hybrid nanostructures have been of central interest in condensed matter physics in the last years since they hold significant promise as platforms for a variety of quantum devices~\cite{Nat_rev_2020, Laubscher_2021, Marra_2022, MF_review_Science_2023}. Such systems can be used to fabricate Cooper pair splitters capable of efficiently generating high-fidelity spatially separated spin-entangled states via the crossed Andreev reflection (CAR) processes~\cite{Choi2000, Loss_2000, Recher_2001, Lesovik2001, Melin_2004, Beckmann_2004, Hofstetter_2011, Schindele_2012, Csonka_2015, Trocha_2015, Trocha_2018, Ranni_2021, Maisi_2022, Andreas_spin_ent_N, Wang_N_2022, Tunable_car_PRL_theory, CPS_2DEG, CPS_dynamics}. Additionally, semiconductor nanowires with strong Rashba spin-orbit interaction (SOI) proximitized by a bulk superconductors offer a potential avenue for realizing synthetic topological superconductors, hosting zero-energy Majorana bound states \cite{Kitaev_org,  Volovik_1999, Read_2000, Ivanov_2001, Nayak_2008, Alicea_2012}. Moreover, by appropriately gating such nanowires, one can create arrays of quantum dots coupled via superconducting sections, forming a platform for realizing fine-tuned versions of topological superconductors, known as minimal Kitaev chains~\cite{Sau_2012_Minimal_Kitaev, poor_mans_MF, Fulga_MKC, MKC_ex, MKC_ex_3dots}, which host ``poor man's" Majorana bound states. Furthermore, superconductor-semiconductor hybrids can be used to create spin qubits in quantum dots that could be manipulated via coupling to superconducting leads ~\cite{Choi2000,Spethmann2022,Spethmann2024} or to create Josephson junctions, which can also host Andreev bound states (ABSs) again suitable for encoding qubits~\cite{Andreev_sq_th, Padurariu_2010, Zazunov_2003, Janvier_2015, Park_2017, Tosi_2019, Cerrillo_2021, Mishra_2021, Andreev_sq_exp, Vidal_2023, Spectr_Spin_Andreec_SC, Hoffman_2024}. Another quantum information related application of such devices is to couple spin qubits hosted in quantum dot over long distances ~\cite{Choi2000,Leijnse_coupling_2013, Hassler_coupling_2015, Rosado_coupling_2021, Spethmann2022, Spethmann2024}.
The possibility of precise and efficient control between the CAR and elastic cotunneling (ECT) processes is crucial for realizing all aforementioned devices.

\begin{figure}[tb!]
\centering
\includegraphics[width=8.6cm]{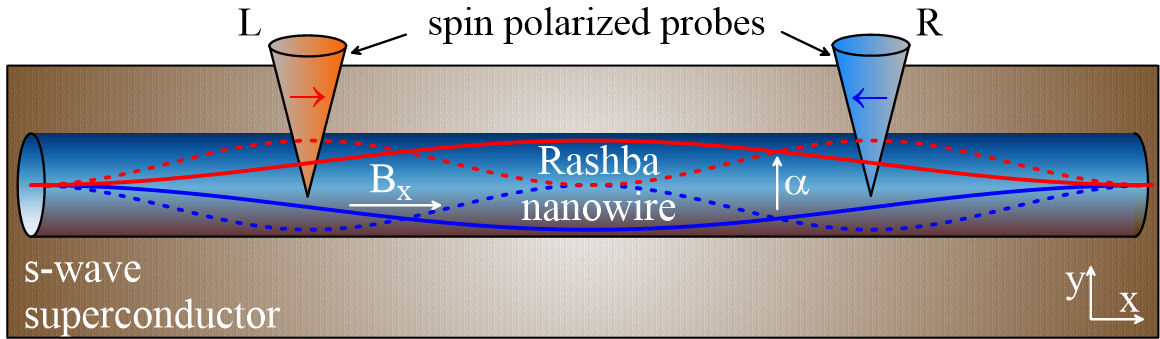}
\caption{Schematics of Rashba nanowire (blue cylinder) proximitized by an $s$-wave superconductor (brown). The magnetic field $B_x$ is applied in  $x$ direction  (nanowire axis) and the  Rashba spin-orbit vector $\alpha$ points in $y$ direction. Such systems can host low-energy spin-down (up) polarized quasiparticle states in the trivial (topological) phase. The energy of these one-dimensional states is limited above by the superconducting gap of the three-dimensional $s$-wave superconductor.
The vertical cones indicate tunnel-coupled side probes (leads or STM tips) that are oppositely spin polarized (blue and red arrows)  and are placed at some distance away from the nanowire ends.}
	\label{Setup}
\end{figure}

Recent state-of-the-art experiments have demonstrated unprecedented control over nanofabrication processes~\cite{Shabani_2016, Zhang2017, Gazibegovic_2017, Thomas_2019, Mayer2020, Heedt_2021, Moehle2021, OConnell_2021, Badawy_2024} and precise tuning of the parameters of such devices~\cite{Mazur_2022, Loo_2023, Tunable_car_PRX}. One of the most accessible experimental methods to investigate and study properties of quasiparticles hosted in such hybrid structures are quantum transport techniques. The local transport spectroscopy techniques can provide some insight about localized in-gap quasiparticle states  e.g zero energy Majorana fermions or non-zero energy ABSs via local conductance peaks as predicted by theory~\cite{Law2009, Akhmerov_2009, Sau_2010, Flensberg2010, Wimmer_2011, Fidkowski_2012, Diego_2013, Chevallier_2016} and tested experimentally~\cite{Sasaki_2011, Mourik2012, Das2012, Deng_2012, Churchill_2013, Deng_2016, Nichele_2017, Lutchyn_2018, Anselmetti_2019, Vaitiekenas_2020, Schneider_2022, Wang_2022}. However, one has to keep in mind that such local transport measurements can be inconclusive in detecting topological superconducting phases since the observed zero-bias peaks can have origins that are different from Majorana fermions~\cite{Kells_2012, Lee_2012, Lee_Nat_Nan_2014, Cayao_2015, Ptok_2017, Liu_2017, Reeg_2018, Penaranda2018, Moore2018, Vuik2019, Woods2019, Liu2019, Chen2019, Pan_2020, Alspaugh2020, Junger2020, Valentini2021, Yu_2021,Hess_2021_sc,  Hess_2023}.
On the other hand, multiterminal transport techniques can provide additional information about the system such as superconducting energy gap anisotropy, nonlocal nature of the quasiparticles, their effective charge associated with electron and hole composition, induced gap closing, and, potentially, about the TPT~\cite{Flatte_1995, Wohlman_2008, Schindele_2014, Schindele_2014, Lobos_2015, Gramich_2017, Rosdahl_2018, Zhang_2019, Nonlocal_BCS_charge_th, Melo_2021, Pikulin_arxiv, Pan_2021, Hess_2021, Menard_2020, Puglia_2021, Poschl_2022, Dourado_arxiv, Hess_2023, topological_gap, Thamm_2024}.

Recently, significant tunability between the CAR and ECT processes has been reported in  nonlocal quantum transport experiments in proximitized Rashba nanowires coupled to quantum dots~\cite{Tunable_car_PRX}. This a key step toward generating fine-tuned topological superconducting phases within the minimal Kitaev chain model~\cite{MKC_ex, MKC_ex_3dots} and generating Bell states in a controllable manner via CAR~\cite{Andreas_spin_ent_N}.

Here, we investigate theoretically spin properties of low-energy quasiparticles hosted in proximitized Rashba nanowires using nonlocal quantum transport simulation techniques. In contrast to previous works, we consider systems with  tunnel-coupled side leads that are also spin polarized. We demonstrate a direct relationship between the sign of the nonlocal conductance and the sign of the quasiparticle spin in systems where at least one side lead is spin polarized. This relationship is observed over a wide range of system parameters. Moreover, since the spin density and its sign can be non-uniform along the nanowire, the ECT and CAR processes, which are correlated with the sign of the nonlocal conductance, depend on the positions at which the side-coupled spin-polarized leads are attached. This  could be used to detect band inversion associated with the sign flip of spin and charge of low-energy quasiparticles~\cite{Spin_and_charge}, which build a low-energy Andreev band. Thus, this property could serve as an additional criterion for verifying topological superconductivity. We note that the energy of these one-dimensional states, forming the Andreev band, is limited from above by the superconducting gap of the three-dimensional parent $s$-wave superconductor. If the quasiparticle energy is above this value, the corresponding state is no longer confined inside the nanowire  and, instead,  gets delocalized over the entire system including the superconductor. Thus, the detection scheme proposed here is only applicable to states within the Andreev band but not to higher energy quasiparticle states.

Moreover, we identify  parameter regimes for which the quasiparticle charge is nearly zero, indicating an equal amount of electron and hole components even though the system undergoes band inversion. In such a regime, the sign of nonlocal conductance related to the quasiparticle spin are particularly robust to onsite disorder. For completeness, we also calculate the local conductance, which also depends on the spin polarization of the probed quasiparticle states, obtaining results consistent with the spin-selective Andreev reflection process \cite{SSAR, SSAR_, SSAR2_ex, SSAR2_th, SSAR_Dom, Jeon_science_2017, Jeon_prb_2018, Wang_prl_2021, Karan_2024}. On the other hand, in setups with two normal leads, we obtain a very good mapping between the sign of nonlocal conductance and the sign of  quasiparticle charge as expected theoretically~\cite{Nonlocal_BCS_charge_th,Tunable_car_PRL_theory} and verified experimentally~\cite{Tunable_car_PRX}.

Our findings describe a general phenomenon and can be used to precisely detect the local spin density of quasiparticles not only in proximitized Rashba nanowires but also in other systems hosting spin-polarized quasiparticle states such as Yu-Shiba-Rusinov states (YSR)~\cite{Yu_1965, Shiba_1968, Rusinov_1969, Yazdani_1997}, spin chains~\cite{Schneider_2022, Choy_2011, Martin_2012, Yazdani_2013, Pientka, Klinovaja_2013, Braunecker_2013, Vazifeh_2013, Yazdani_2014, Ruby_2015, Pawlak_2016, Kuster_2022}, Andreev spin qubits~\cite{Andreev_sq_th, Park_2017, Mishra_2021, Andreev_sq_exp, Vidal_2023, Spectr_Spin_Andreec_SC, Hoffman_2024}, Caroli-de Gennes-Matricorn and Majorana vortex states~\cite{Volovik_1999, SSAR2_ex, Caroli1964, Xu2015, Chiu2020} etc. Furthermore, our results offer insights for designing devices functioning as Cooper pair splitters and for realizing poor man's Majorana fermions in minimal Kitaev chain models in which precise tunability over ECT and CAR processes is crucial~\cite{Tunable_car_PRX,MKC_ex, MKC_ex_3dots}.
In our simulations, we employ realistic parameters. It is noteworthy that the coupling of side-leads to nanowires has already been experimentally demonstrated,  both for normal leads~\cite{topological_gap}, for quantum dots~\cite{Lee_Nat_Nan_2014,spin_filter_delftNC,spin_filter_dots_Marcus}, and, notably, for ferromagnetic leads~\cite{Andreas_spin_ent_N}. Hence, our predictions are directly amenable to experimental verification.

\section{Model}
As an example of a system that can host spin polarized low-energy one-dimensional Andreev band with quasiparticle wavefunctions extended over the entire length of the nanowire, we consider a Rashba nanowire proximitized by a three-dimensional  $s$-wave superconductor. It was shown that low-energy quasiparticle states in such systems for certain parameter ranges have a well-defined spin polarization~\cite{Spin_and_charge}.
To be specific, we consider a one-dimensional Rashba nanowire aligned along the $x$-axis and placed on top of an $s$-wave superconductor in the presence of an external magnetic field applied along the nanowire axis (see Fig.~\ref{Setup}). The system can be modeled by the tight-binding Hamiltonian:
\begin{align}
&{H}=\sum_{j=0}^{N-1}[{\Psi}_{j+1}^\dag(-t\tau_z\sigma_0-i\tilde{\alpha}\tau_z\sigma_y){\Psi}_j+{\text {H.c.}}]\nonumber\\
&\hspace{10pt}+\sum_{j=0}^N{\Psi}_j^\dag[{(2t -\mu)}\tau_z\sigma_0+\Delta_{sc}\tau_y\sigma_y+\Delta_{Z}\tau_z\sigma_x]{\Psi}_j \label{hamiltonian_position},
\end{align}
where ${\Psi}_j=({c}_{j \uparrow}, {c}_{j \downarrow}, {{c}^{\dag}_{j \uparrow}}, {{c}^{\dag}_{j \downarrow}})^T$ is given in standard Nambu representation.  The creation operator ${c}^{\dag}_{j \sigma} $ acts on an electron with spin $\sigma$ located at site $j$ in a chain of $N$ sites with lattice constant $a$. The Zeeman energy $\Delta_Z=g\mu_B B_x/2$ is determined by the strength of the external magnetic field applied along the $x$-axis, $B_x$, and by the $g$-factor. The proximity effect by the $s$-wave superconductor is responsible for inducing a uniform superconducting pairing term  $\Delta_{sc}$ in the nanowire.  The chemical potential of the nanowire $\mu$ is calculated from the SOI energy  and  $t={\hbar}^2/(2m^{*}a^2)$ is the hopping amplitude, where $m^*$ is the effective mass and $a$  the lattice spacing used in the effective tight binding modeling. The Pauli matrices $\sigma_i$ ($\tau_i$) act on spin (particle-hole) space and $\tilde{\alpha}=\alpha/(2a)$ is the spin-flip hopping amplitude resulting from the Rashba SOI characterized by the strength $\alpha$ and $E_{\rm SO}=\tilde{\alpha}^2/t$ is the associated SOI energy. In order to find the energy spectrum $E_n$ and corresponding wavefunctions $\Phi_n (j)$ labeled by the index $n=1,...,4N$ we diagonalize the Hamiltonian ${H}$ numerically. The quasiparticle spin and charge density distribution for given energy eigenstates $E_n$ are defined, respectively, as follows:
\begin{align}
&S_x(j, E_n)=\Phi_n^\dag(j)\tau_z\sigma_x\Phi_n(j)\\
&Q(j, E_n)=\Phi_n^\dag(j)\tau_z\sigma_0\Phi_n(j)
\end{align}
Below, we show that the transport properties of the investigated system strongly depends on these quantities.

In order to investigate the transport properties of the system we calculate numerically  matrix elements of the differential-conductance $G_{ij}=dI_i/dV_j$ defined  as derivative of the total current $I_i$  flowing into the nanowire from the lead $i$ with respect to the voltage bias $V_j$ applied to the lead $j$ as described in detail in Refs.~\cite{Nonlocal_BCS_charge_th,Hess_2021}, while the superconductor is grounded. To be specific we employ Blonder-Tinkham-Klapwijk formalism~\cite{BTK} in which the zero-temperature nonlocal conductance $G_{ij}(E)$ between leads $i$ and $j$ is given by
\begin{equation}
G_{ij}(E)=\frac{e^2}{h}[T_{ij}^{ee}(E)-T_{ij}^{he}(E)]
\end{equation}
and the local conductance by
\begin{equation}
G_{ii}(E)=\frac{e^2}{h}[N_i-R_{ii}^{ee}(E)+R_{ii}^{he}(E)],
\end{equation}
where $T_{ij}^{ee}(E)$ [$R_{ii}^{ee}(E)$] and $T_{ij}^{he}(E)$ [$R_{ii}^{he}(E)$] are, respectively, the amplitudes of the normal (electron-electron) and Andreev (electron-hole) transmission [reflection] process for the charge carriers with energy $E$ injected in the system from $j$-th lead and transmitted towards the $i$-th lead, where  $i,j=L,R$ label the left and right leads. The number of modes in the leads is denoted by $N_{i}$. For the system considered here, the conductance $G_{ij}(E)$ is calculated numerically using the Kwant~\cite{kwant} package and Adaptive~\cite{adaptive} for the optimal parameter sampling. In all displayed results, the conductance is expressed in units of $e^2/h$. In our simulations, the leads are modeled by the Hamiltonian $H_{lead}(k)~=~(2t[1~-~\cos(ka)]~-~\mu_{\rm lead})\tau_z\sigma_0~+~ M_x\tau_z\sigma_x$, 
which for $M_x=0$ describes the normal (unpolarized) lead and for $M_x<0$ ($M_x>0$)  the corresponding  spin up (down) polarized  leads.

\begin{figure}[!tb]
	\centering
	\includegraphics[width=8.6cm]{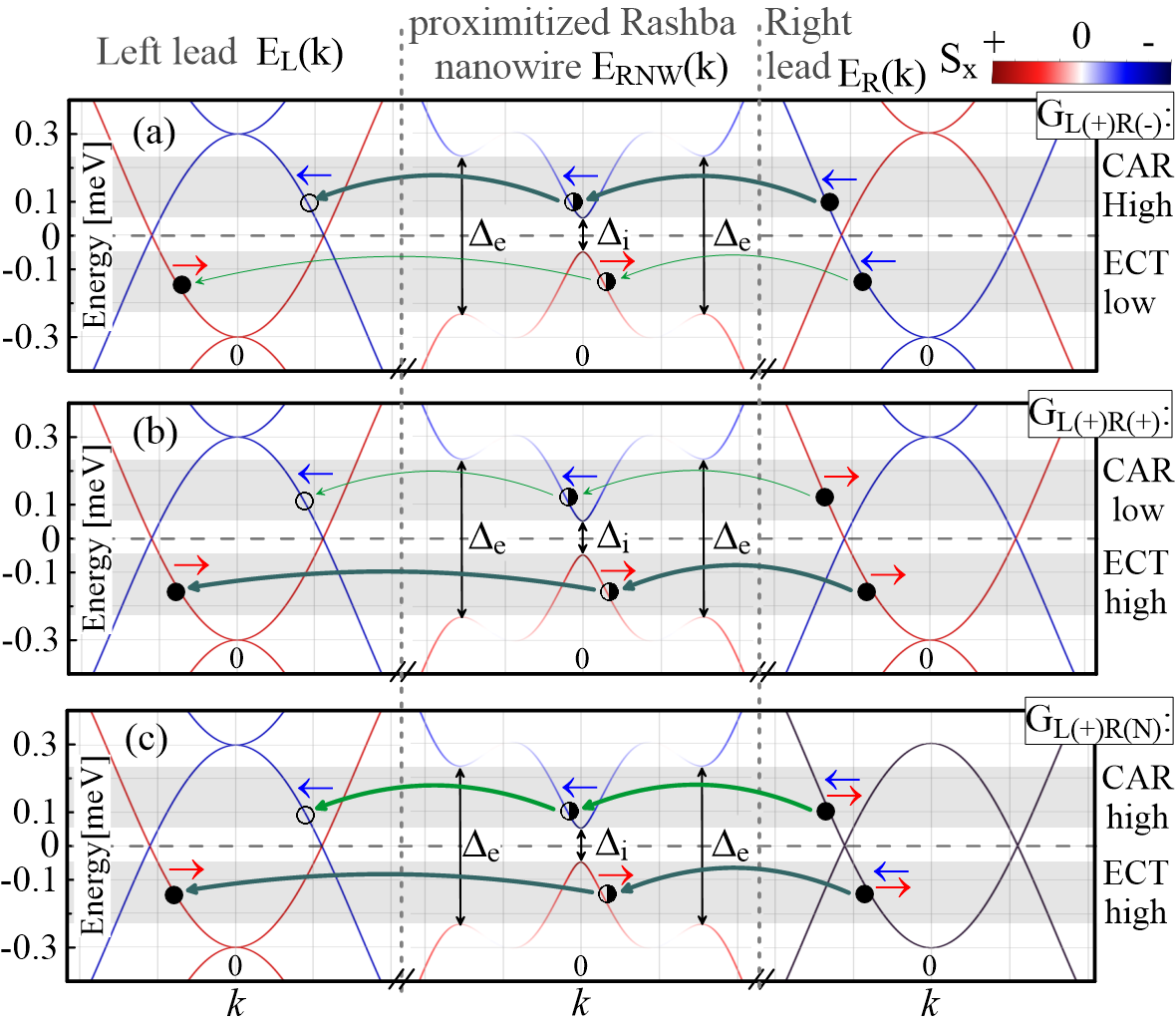}
	\caption{Energy bands of the left spin up polarized  lead $E_L(k)$, the proximitized Rashba nanowire $E_{RNW}(k)$ and the right lead which is either (a) spin-down or (b) spin-up polarized or (c) normal $E_R(k)$. The $x$ component (along B-field) of the spin polarization of quasiparticles from the Andreev band, $S_x(k)$, is indicated by the colorbar. Shaded gray areas mark an energy window $\Delta_i<E<\Delta_{ex}$ in which spin sensitive transport occurs. Among many possible scattering processes, we denote the dominant ones contributing to the nonlocal conductances (a) $G_{L(+)R(-)}$,   (b) $G_{L(+)R(+)}$, and (c) $G_{L(+)R(N)}$, which are either CAR (negative conductance) or ECT (positive conductance) processes. The strength of the corresponding signal (high or low) is reflected in the line thickness. Empty circles denote holes while filled ones denote electrons and half-filled ones correspond to quasiparticles in the nanowire. We set the following parameters for the nanowire: $\mu=0$, $t=0.27$ meV,  $\Delta_{sc}=0.25$~meV, $\Delta_Z=0.2$~meV (nontopological phase), $\alpha=50$ meVnm ($E_{\rm SO}=0.23$ meV). For the spin up (down) polarized leads, we set $\mu_{\rm lead}=0$ and $M_x=-0.3$ meV ($M_x=0.3$ meV) and, for the normal lead, we set $\mu_{\rm lead}=0.3$ meV and $M_x=0$. }
	\label{bands}
\end{figure}

The corresponding band structure of the leads $E_{L,R}(k)$ as a function of momentum $k$ is schematically depicted in Fig.~\ref{bands} together with the band structure of the proximitized Rashba nanowire $E_{RNW}(k)$ and quasiparticle spin $S_x(k)=\Phi^\dag(k)\tau_z\sigma_x\Phi(k)$ (see App. A for details). Again, the spin quantization axis is assumed to be in the  $x$ direction, along the applied magnetic field. The leads are tunnel-coupled to the nanowire at the positions $x_L$ and $x_R$, respectively, for the left and the right lead with positive hopping amplitude $t_\Gamma (<t)$. 

For the purpose of this study, we choose system parameters that are within  experimental reach and for regimes where one can get a substantial number of extended quasiparticle states forming the one-dimensional Andreev band inside the superconducting gap of the three-dimensional $s$-wave superconductor (long nanowire limit $L=10$~$\mu$m). To be specific, we choose the  following parameters for Rashba nanowires:  effective mass $m^*=0.014m_0$, $g$-factor $g=50$, $\Delta_{sc}=0.25$~meV, and $\alpha=50$~meVnm ($E_{\rm SO}=0.23$~meV). In case for spin up (down) polarized leads $\mu_{\rm lead}=0$ and $M_x=-0.3$~meV ($M_x=0.3$~meV) and for the normal lead $\mu_{\rm lead}=0.3$~meV and $M_x=0$. For the purpose of numerical efficiency we set $a=100$~nm corresponding to $t\approx0.27$ meV and $N=100$.  However, we have checked that for $a=10$~nm ($t\approx27$ meV) and $N=1000$ the key results are very similar (see App.~C for details). To study the dependence on the Zeeman energy, we set  $\mu=0$, and vary $\Delta_Z$ between 0 and $2\Delta_{sc}=0.5$~meV. To study the dependence on the chemical potential $\mu$, we fix the Zeeman energy to $\Delta_Z$=0.4~meV and change $\mu$ between -0.6 and 0.6 meV. In most  cases the leads are attached symmetrically around the nanowire center at the positions $x_L=(1/4)L$ and $x_R=(3/4)L$, unless stated otherwise. However, as shown for disordered systems and in  App.~B, as long as the leads are attached sufficiently far away from the nanowire ends the main features of the spin-dependent nonlocal conductance are not affected.
Here, we work in the tunneling regime with $t_\Gamma<t$, so individual quasiparticle states can be resolved in transport simulations and compared with the local spin and charge densities of the quasiparticels from the Andreev band obtained from finite-size calculations.  Furthermore, the weak coupling regime between the spin polarized lead and the nanowire can be advantageous in suppressing its potentially diminishing effects on superconductivity in  case of spin polarized leads. Here, we choose parameters for InAs in order to demonstrate that the conductance inversion (spin-dependent nonlocal conductance)  can be observed for realistic parameter values. However, we would like to emphasize that the  spin-dependent 
behavior--switching of the conductance sign--has a rather universal character and can be used for  quasiparticle spin detection in other systems and for other parameter regimes.

\section{Results}
\subsection{Schematic picture of nonlocal transport}

We present first a schematic (physical) picture of the dominant processes contributing to the nonlocal conductance $G_{LR}$ for three different configurations of the leads, which is then followed by the presentation of exact numerical results in subsequent sections. Regarding the proximitized Rashba nanowire, the extended Andreev band quasiparticles with energies $E_{RNW}(k)<\Delta_{sc}$ have a certain sign of the spin if the exterior gap exceeds the interior one, $\Delta_{i}<\Delta_{ex}$, which for $\mu=0$ is satisfied if $0<\Delta_{Z}<2\Delta_{sc}$. In the topologically trivial state (here, $\Delta_Z<\Delta_{sc}$ for $\mu=0$), the quasiparticle states with the lowest positive energy around $k\approx0$ have negative spin polarization $S_x(k)$  as illustrated in Fig.~\ref{bands}(a)-(c) (middle panels). 

When the left lead is spin-up (+) and the right lead spin-down polarized (-), the spin of the injected charge carrier with positive energy matches the negative spin of quasiparticles in the proximitized region.  However, due to spin mismatch it cannot enter the left lead as an electron and has to be transformed into hole which is spin-down polarized [see Fig.~\ref{bands}(a) for details]. This leads to a strong CAR dominated nonlocal conductance $G_{L(+)R(-)}$ signal characterized by a negative sign. 

Next, we consider the case with two spin-up polarized leads. Here, even though the injected spin-up polarized charge carrier with positive energy does not match the spin direction of the spin-down quasiparticle in the proximitized region, transport can still occur due to the presence of SOI, however, with low probability transfer amplitude  giving rise to a small CAR signal. However, when we look  at negative energies, the spin polarizations of electrons  in both leads and of the quasiparticle in the nanowire are the same. Thus, the  injected electron can easily enter the proximitized region and leave it as an electron with the same spin. In such a scenario nonlocal transport is dominated by strong ECT  as depicted in Fig.~\ref{bands}(b) with a positive sign of the nonlocal conductance $G_{L(+)R(+)}$. 

Finally, we consider the case with the left lead being spin-up polarized while the right lead is normal (unpolarized). Due to spin degeneracy in the right lead, the electron can freely enter the proximitized region regardless of the quasiparticle spin polarization and enter the left lead as spin-down (up) hole (electron). This supports strong CAR (ECT) dominated nonlocal conductances for positive (negative) energies [see Fig.~\ref{bands}(c) for details]. 
In the following sections, we support the picture presented here by exact numerical simulations of  the spin-dependent quantum transport.

\subsection{Quasiparticle spin detection}
We next study numerically the low-energy states and their spin and charge densities for Rashba nanowires and the corresponding local and nonlocal conductances. In these calculations, we tune either the Zeeman energy $\Delta_Z$ or the chemical potential $\mu$ symmetrically around given critical values corresponding to the TPT. To study the spin-dependent transport, like before, we consider different configurations of the leads attached to the system: i) leads with the same spin polarization (either up or down), ii) leads with opposite spin polarization, and iii) a setup where the left lead is spin polarized while the right lead is normal (unpolarized).

First, we calculate numerically the energy spectrum for  finite-size nanowires as  function of $\Delta_Z$ together with the local quasiparticle spin $S_x(j, E_n)$ [see Fig.~\ref{Fig_Gii_Gij_NSN_FSF_Dz}(a)] for selected positions $x_L$ or $x_R$ at which the leads will be attached in the  transport simulations. This will allow us to compare and relate the sign of the local quasiparticle spin with the sign of the nonlocal conductance. We note that for the uniform system without disorder the spin density profile of selected bulk states are symmetric with respect to the center of the nanowire: $S_x(x)=S_x(L-x)$  as presented in Fig.~\ref{Fig_Gii_Gij_NSN_FSF_Dz}(d). We observe that the quasiparticle spin $S_x(x_L)$ of the low-energy quasiparticles  at position $x_L$ (which should be sufficiently far away from the nanowire end) changes its sign when the system undergoes the TPT as  the Zeeman energy passes the critical value $\Delta_Z^c=\Delta_{sc}$ for $\mu=0$. 

We start the discussion with presenting results from transport simulations for systems with two leads that have the same spin polarization either up (+) or down (-). We observe a strong  positive nonlocal conductance $G_{L(\pm)R(\pm)}(\Delta_Z,E_F)=G_{R(\pm)L(\pm)}(\Delta_Z,E_F)$, if the spin of the probed ABS is the same as the spin of the electrons in the leads. In this case the transport is dominated by  ECT processes. For example, when the leads are spin up
\onecolumngrid
\begin{center}
	\begin{figure}[H]
		\centering
		\includegraphics[width=17.5cm]{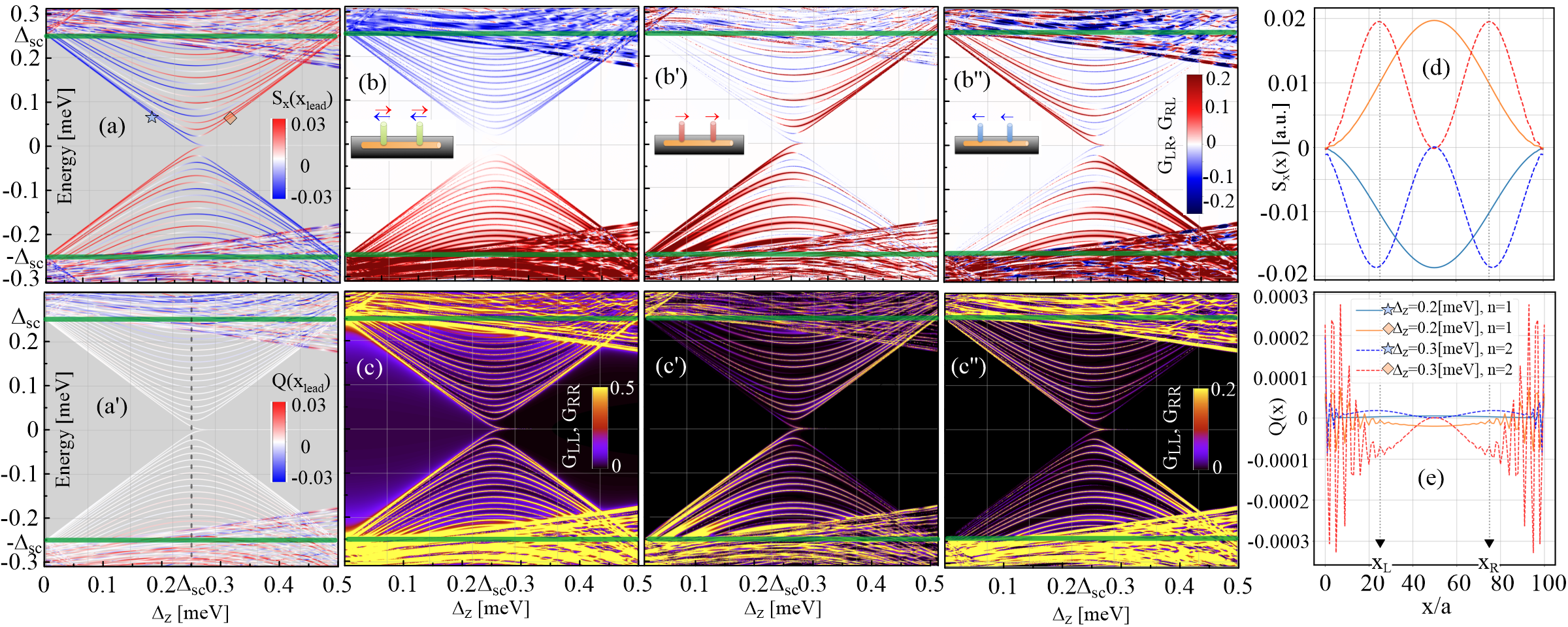}
		\centering
		\caption{ (a) Energy spectrum of proximitized Rashba nanowire as a function of Zeeman energy $\Delta_Z$ induced by the magnetic field $B_x$ applied along the nanowire axis, which also defines the spin quantization axis. The color bar represents  the $x$-component of the spin density, $S_x(j,E_n)$, for a given energy eigenstate $E_n$ at position $j=x_L/a$ and $x_R/a$ with $x_L/a=25$ or $x_R/a=75$ (here the spin density is symmetric with respect to the nanowire center). On panel (a') the color bar represents the quasiparticle charge $Q(j,E_n)$. The vertical dashed line indicates the critical value of $\Delta_Z^c=\Delta_{sc}$ at which the system undergoes a gap closing and reopening at the TPT  point characterized by spin inversion of the lowest-energy states. The nonlocal conductance $G_{LR}(\Delta_{Z}, E)=G_{RL}(\Delta_{Z}, E)$ for systems with (b) normal - spin unpolarized (degenerate) leads, (b') both spin-up  
			and (b'') spin-down 
			polarized leads. The colored arrows on the insets denote the spin polarization of the leads. The corresponding local conductance $G_{LL,RR}(\Delta_{Z}, E)$ is plotted on the panels (c)-(c'').  The corresponding strength of local conductance signals (c', c'') is consistent with the spin-selective Andreev reflection. Quasiparticle  (d) spin $S_x(x,E_n)$ and (e) charge  density for the two lowest nonzero energy ABSs in the trival ($\Delta_Z$=0.2 meV) and topological ($\Delta_Z$=0.3 meV) phase with  negative and positive spin density, respectively, and almost zero charge density in both cases. On panels (b', b''), one can observe strong positive (weak negative) nonlocal conductance signal related with dominant ECT (CAR) process when the spin of the leads matches (are opposite to) local spin polarization of ABS.  In contrast to spin, we note that we do not observe any correlations between (b) nonlocal conductance and (a') quasiparticle charge for the system with normal leads as the quasiparticle charge is almost zero (a', e). The  nanowire-lead coupling is set to $t_\Gamma=0.4t\approx0.1$ meV while the rest of the parameters are as in Fig.~\ref{bands}.}
		\label{Fig_Gii_Gij_NSN_FSF_Dz}
	\end{figure}
	\begin{figure}[H]
		\centering
		\includegraphics[width=17.5cm]{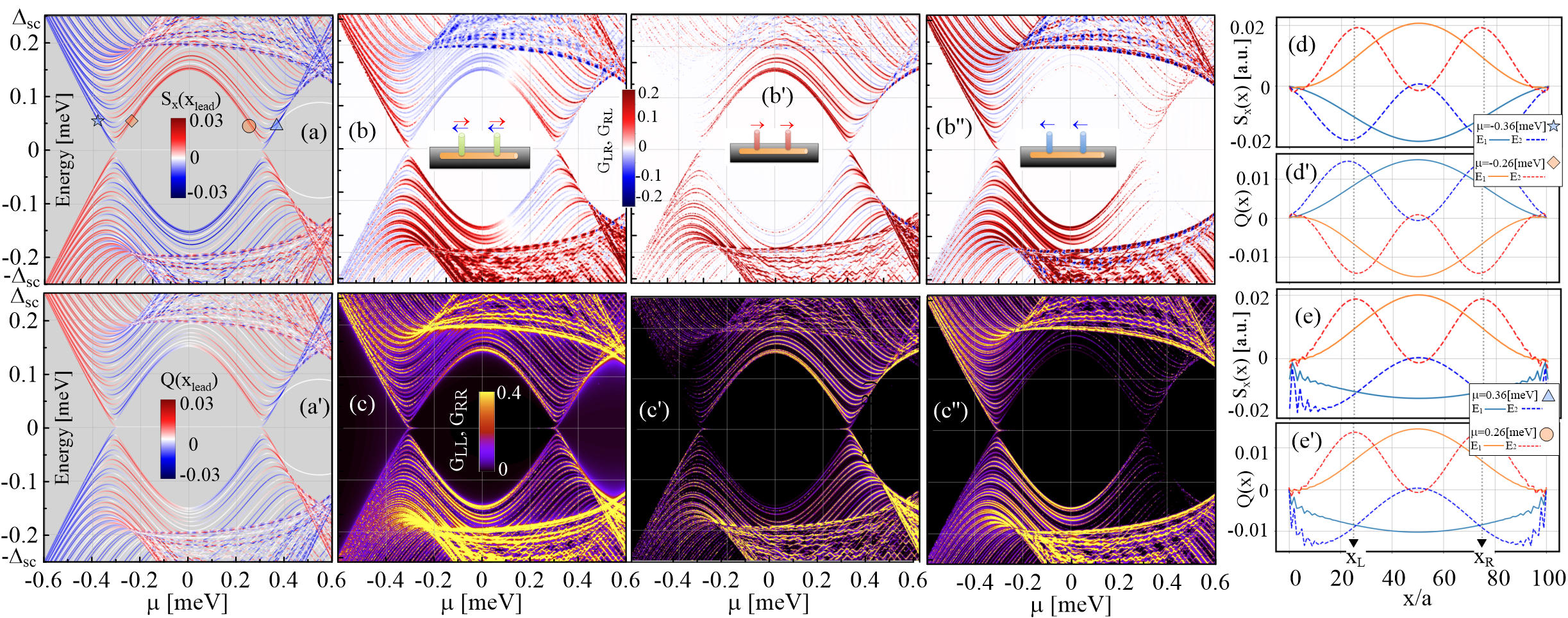}
		\caption{Same as for Fig. \ref{Fig_Gii_Gij_NSN_FSF_Dz} but now as a function of chemical potential $\mu$ while the Zeeman energy is set to $\Delta_Z=0.4$~meV. (a') For a wide range of $\mu$, one can clearly see nonzero values of the quasiparticle charge. As a consequence, the sign of the nonlocal conductance $G_{L(N)R(N)}(\mu, E)$ for systems with normal leads in panel (b) corresponds quite well to the quasiparticle charge shown in  panel (a').  Again, there is a strong positive nonlocal conductance signal for the case when the spin of the probed ABSs is the same as the spin polarization of the leads.}
		\label{Fig_mu}
	\end{figure}
\end{center}
\twocolumngrid

\begin{figure}[!ht] 
	\centering
	\includegraphics[width=8.6cm]{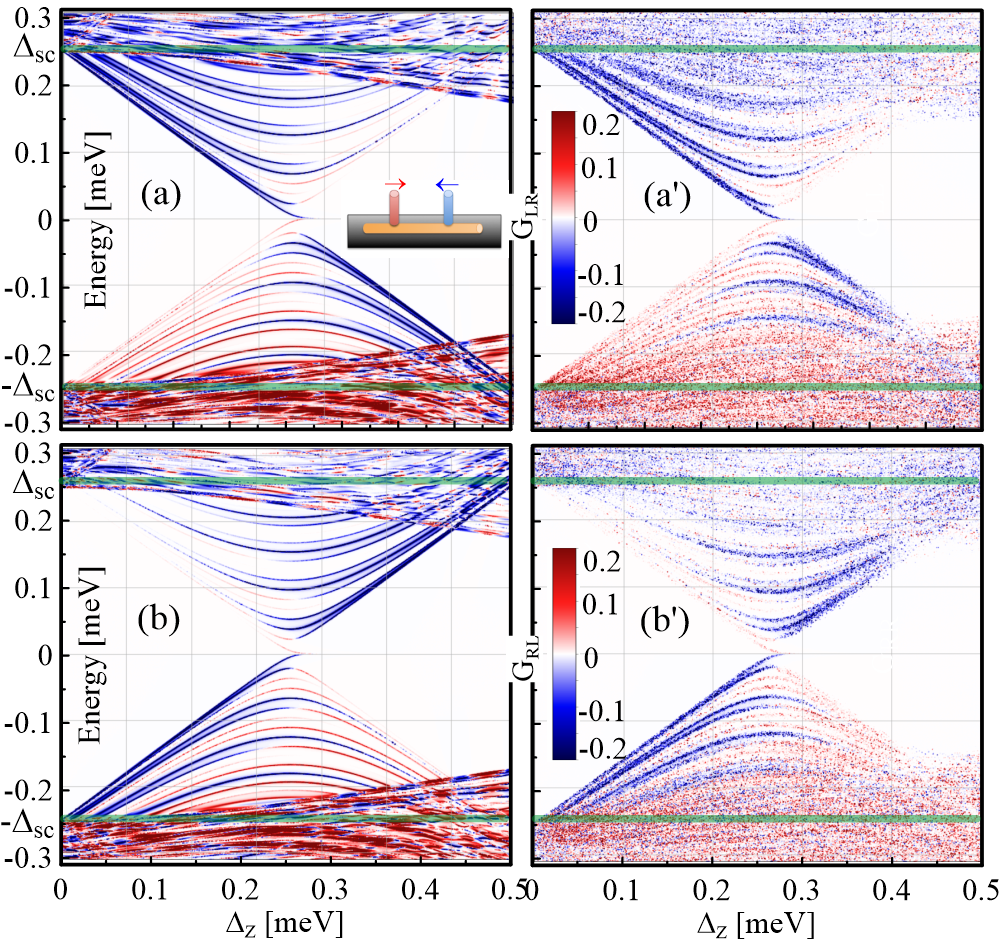}
	\caption{Nonlocal conductance maps (a) $G_{L(+)R(-)}(\Delta_{Z}, E)$ and (b) $G_{R(+)L(-)}(\Delta_{Z}, E)$  for the same parameters as in Fig.~ \ref{Fig_Gii_Gij_NSN_FSF_Dz} but here leads have opposite spin polarizations as indicated in the inset to the panel (a) for the clean system: the left (right) lead is spin-up (down) spin polarized. When the spin of injected charge carriers from the lead is the same as (opposite to) the spin polarization of the probed ABS, there is a strong negative (weak positive) nonlocal  conductance. This means that now nonlocal transport is dominated by CAR process. This behavior can be used to tune the amplitude of CAR processes, being essential for generating entangled states via  Cooper pair splitting processes. Results for systems with moderate disorder $|\delta\mu_j|\le\Delta_{sc}$ are depicted on panels (a', b') and show that the observed behavior of spin-dependent conductance is robust against disorder.}
	\label{Fig_Gij_pFSmF_Dz_nodis_dis}
\end{figure}
\begin{figure}[!ht]  
	\centering
	\includegraphics[width=8.6cm]{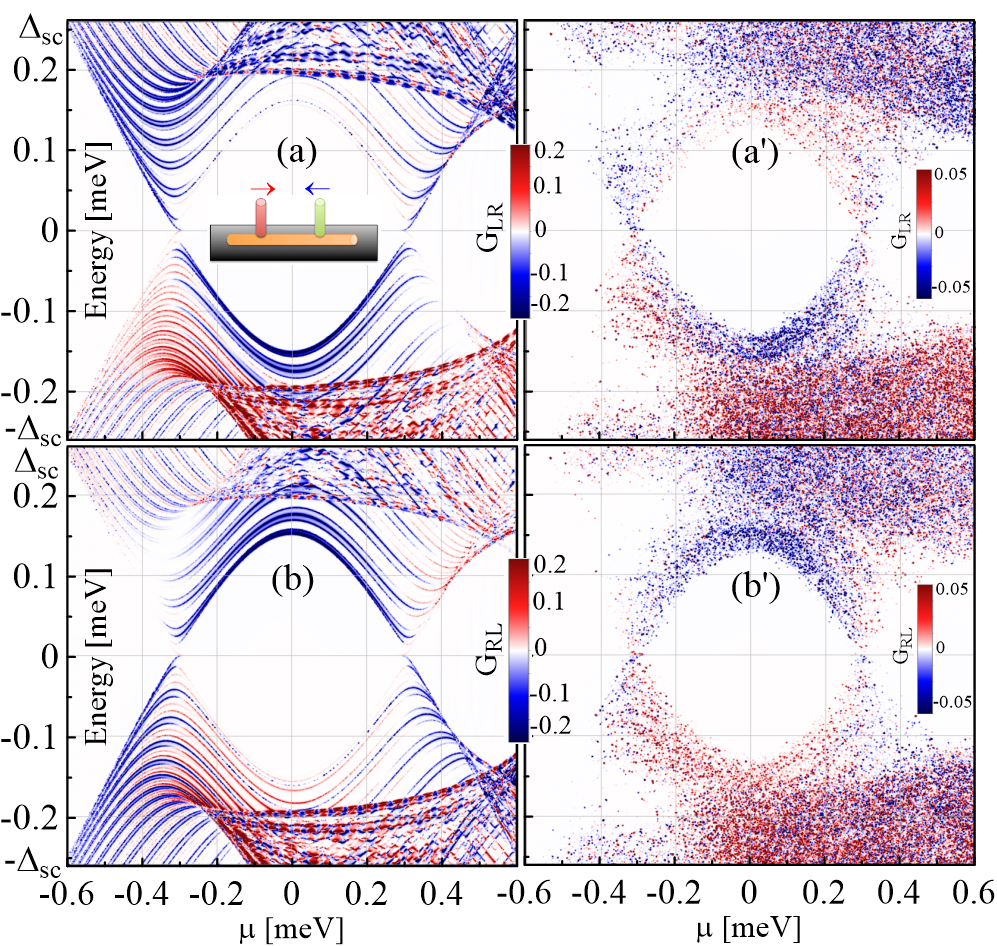}
	\caption{The same as for Fig.~\ref{Fig_Gij_pFSmF_Dz_nodis_dis} but now the chemical potential $\mu$  is varied and the Zeeman energy is kept fixed to $\Delta_Z=0.4$~meV.  Here, we can see again that the transport is dominated by  CAR processes (negative conductance), which can be controlled by tuning $\mu$ (e.g. via backgate).}
	\label{Fig_Gij_pFSmF_mu_nodis_dis}
\end{figure}
\noindent
(spin  down) polarized one can observe strong positive nonlocal conductance signals as depicted in Fig.~\ref{Fig_Gii_Gij_NSN_FSF_Dz}(b')~[Fig. \ref{Fig_Gii_Gij_NSN_FSF_Dz}(b'')] for energies corresponding to the lowest bulk states and values of the Zeeman energies for which corresponding quasiparticle states also  have positive (negative) sign of the local spin density at the positions of the leads. The appearance of weak negative nonlocal conductance signals (CAR) is due to the presence of SOI in the nanowire that allows for normally forbidden spin-flip transport processes. On the other hand, by analyzing plots  of the local conductance shown in Fig. ~\ref{Fig_Gii_Gij_NSN_FSF_Dz}(c') and (c''), we notice that  conductance peaks are more (less) pronounced when the spin polarization in the leads matches (does not match) the spin polarization of the probed quasiparticles. This observation is consistent with so-called SSAR processes~\cite{SSAR, SSAR_, SSAR2_ex, SSAR2_th, SSAR_Dom, Jeon_science_2017, Jeon_prb_2018, Wang_prl_2021}.  We obtain analogous results with ECT dominated signals when the chemical potential   deviates from $\mu=0$ while $\Delta_Z$ is fixed (see Fig.~\ref{Fig_mu}). In this case, there are two TPTs at $\mu_c^{\pm} = \pm \sqrt{\Delta_{Z}^2 - \Delta_{sc}^2}$. Still, we again can identify these TPTs by simply looking at the sign of the nonlocal conductances.

Next, we study setups in which the left and right leads have opposite spin polarizations. Strong nonlocal conductance signals $G_{L(+)R(-)}(\Delta_Z,E)$ are observed in  Fig.~\ref{Fig_Gij_pFSmF_Dz_nodis_dis}(a) and $G_{L(+)R(-)}(\mu,E)$ in Fig.~\ref{Fig_Gij_pFSmF_mu_nodis_dis}(a) [$G_{R(-)L(+)}(\Delta_Z,E)$ in Fig.~\ref{Fig_Gij_pFSmF_Dz_nodis_dis}(b) and $G_{R(-)L(+)}(\mu,E)$ in Fig.~\ref{Fig_Gij_pFSmF_mu_nodis_dis}(b)] when the spin of the injected carriers in the right (left) lead matches the local spin polarization of ABSs in the nanowire at a given bias. We note that ${\rm sgn} [G_{LR}(\Delta_Z,E)]= - {\rm sgn} [G_{RL}(\Delta_Z,E)]$. In such a setup, the nonlocal transport is dominated by CAR processes, a fact that can be advantageous for realization of Cooper pair splitters. The nonlocal conductance has a small positive value when the injected carriers have  spin opposite to the probed ABS states, which is possible, again, due to SOI in the proximitized nanowire. 
\begin{figure}[!t] 
	\centering
	\includegraphics[width=8.6cm]{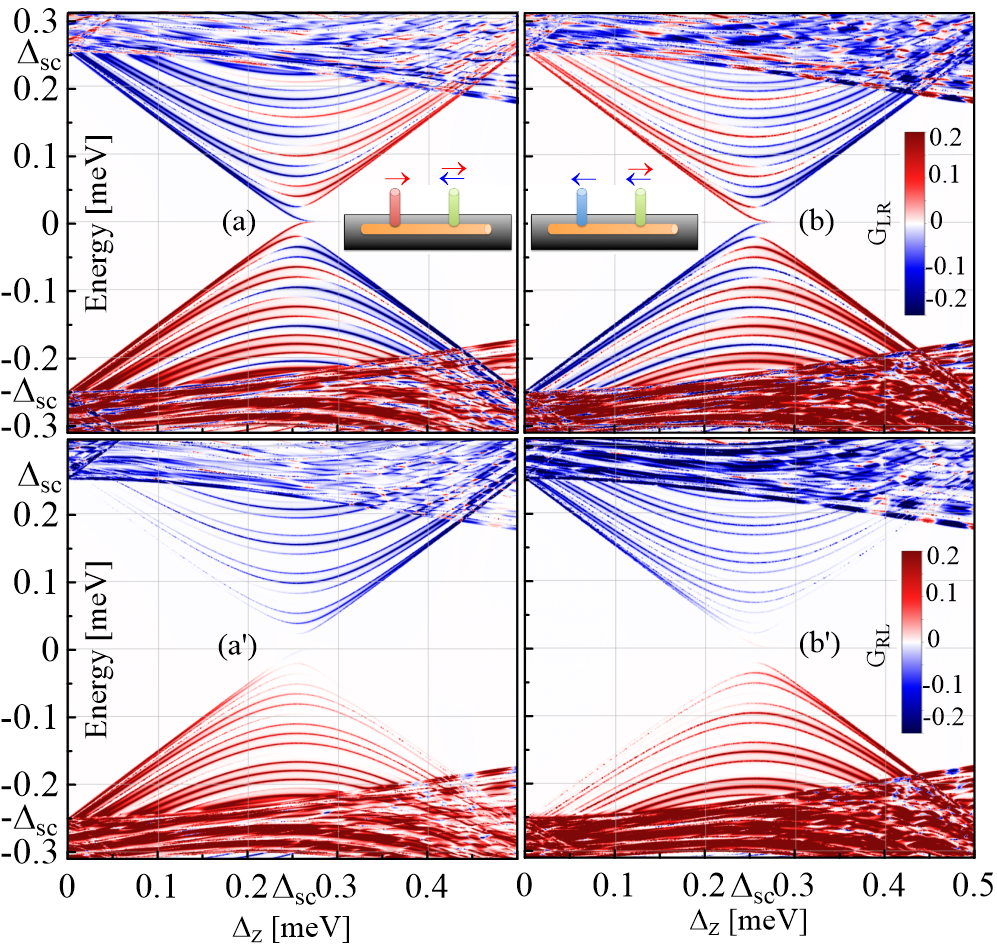}
	\caption{Nonlocal conductance maps $G_{L(\pm)R(N)}(\Delta_Z, E)$ for systems where  right lead is normal (spin unpolarized) and  left lead is either (a) spin-up or (b) spin-down polarized as indicated on insets. Here, we observe the exact (anti) correspondence between the sign of the local spin density of a given quasiparticle and nonlocal conductance $G_{L(\pm)R(N)}(\Delta_{Z}, E)$. When the left lead has same (opposite) spin polarization as the probed quasiparticle, the nonlocal conductance $G_{L(\pm)R(N)}(\Delta_{Z}, E)$ has a large positive (negative) value. For this case, the conductance signal is strongest and such a setup would be most optimal for probing quasiparticle spin polarization via nonlocal transport and tuning between ECT and CAR behavior. For completeness, we plot $G_{R(N)L(\pm)}(\Delta_Z, E)$ - the carriers are injected from spin-up and spin-down polarized leads, respectively, on panels (a') and (b'). The signal is much weaker when the spin polarization of the quasiparticles is opposite to that in the spin-polarized lead.}
	\label{Fig_Gij_pFSN}
\end{figure}
\begin{figure}[!t] 
	\centering
	\includegraphics[width=8.6cm]{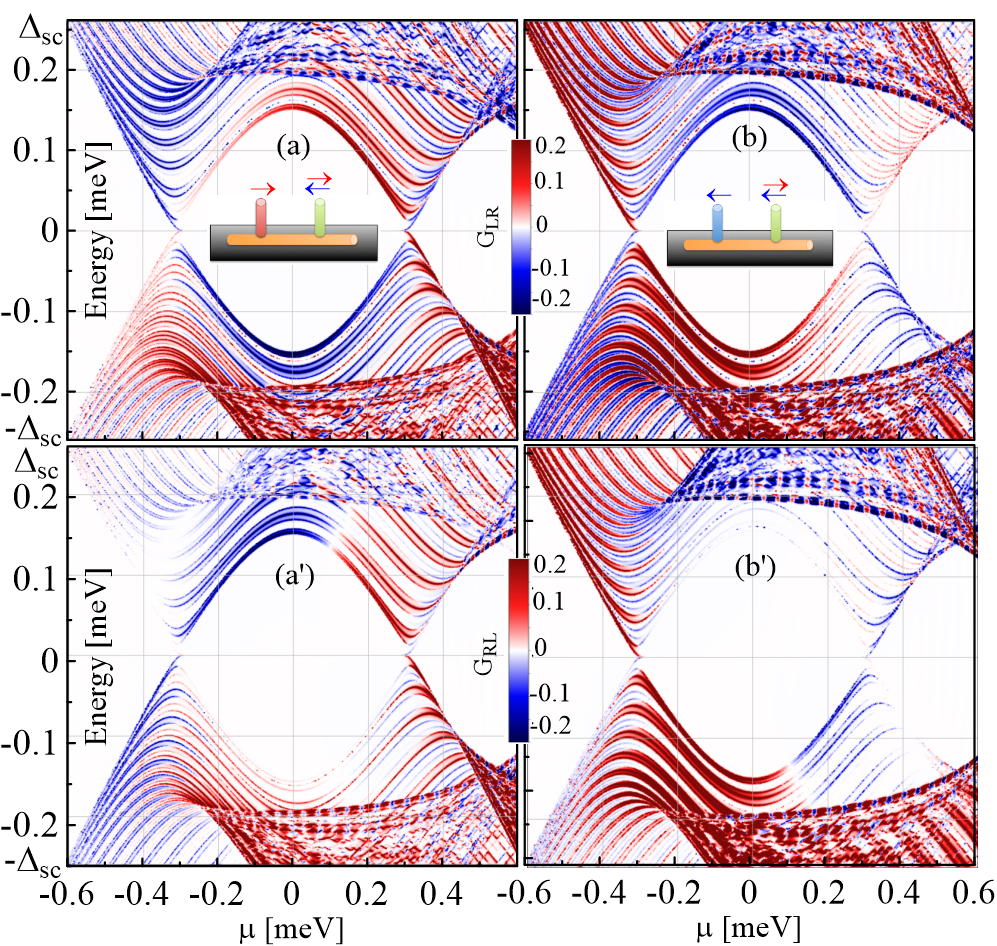}
	\caption{The same as in Fig.~\ref{Fig_Gij_pFSN} but now as a function of chemical potential $\mu$ while the Zeeman energy is set to $\Delta_Z=0.4$ {meV}. One can  again  observe the perfect correspondence between the spin polarization of the probed quasiparticles and the sign of the nonlocal conductance $G_{LR}(\mu, E)$. Here, tunning between the CAR and ECT dominating regimes can be realized by changing the chemical potential.}
	\label{Fig_Gij_FSN_mu}
\end{figure}
\begin{figure}[!h] 
	\centering
	\includegraphics[width=7.25cm]{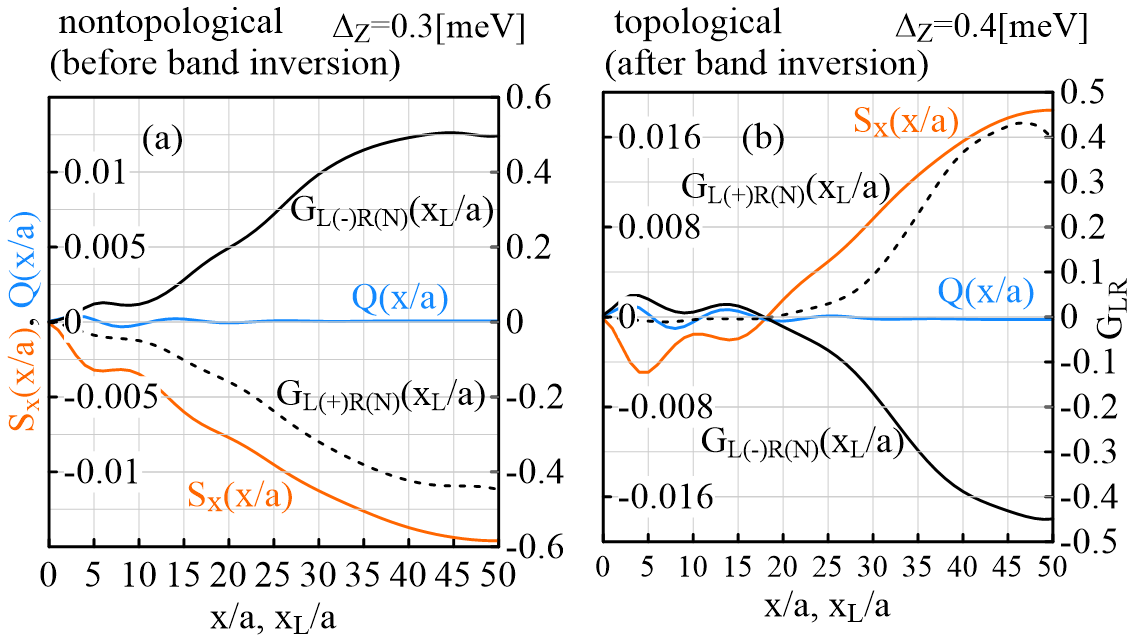}
	\caption{Nonlocal conductance $G_{L(\pm)R(N)}$ for the system with the left lead being spin-up/down ($\pm$) polarized and the right lead being normal (N) as  function of position of the left lead which changes from $x_L=0$ to the nanowire center $x_L/a=L/2/a$, while  the right lead is fixed to $x_R/a=55$. The energy of the injected charge carries matches that of the lowest quasiparticles from the Andreev band (smallest positive nonzero energy). Here,  $Q(x/a)$ and $S_x(x/a)$ denote quasiparticle charge and spin density profile, respectively,  for the system (a) in the nontopological phase for $\Delta_Z=0.3$~meV and (b) in the topological phase for $\Delta_Z=0.4$~meV, while the TPT occurs at  $\Delta_Z^c=\Delta_{sc}=0.35$~meV. The  charge density is almost zero, while the spin density is negative in the nontopological phase and positive in topological one. Here, we show that the sign of the nonlocal conductance $G_{L(\pm)R(N)}$ perfectly matches with  the sign and magnitude of the spin density at a given point. Here, $L=3$~$\mu$m, $a=30$~nm($t\approx3$ meV), $N=100$, $\alpha=40$ meVnm ($E_{\rm SO}=0.15$~meV), $t_\Gamma=0.15t\approx0.45$~meV.}
	\label{Sx_Q_GLR_3}
\end{figure}
We have also considered the case when the left lead is spin polarized and the right one is normal - spin unpolarized (spin degenerate). In such a setup, the nonlocal conductance  $G_{L(\pm)R(N)}$ has a strong positive (negative) value when the polarization of the left lead is the same as (opposite to) the spin polarization of the low-energy quasiparticles from the Andreev band. Such gate configurations make the system an ideal platform for detecting the sign of the local spin polarization of quasiparticles, which is directly linked to the sign of the nonlocal conductance. 
Therefore, such setups are very suitable for the detection of spin inversion of the one-dimensional Andreev bands induced by the TPT. Furthermore, such  setups are optimal for switching between ECT and CAR processes in Rashba nanowires either by tuning the magnetic field [see Fig.~\ref{Fig_Gij_pFSN}(a,b)]  or the chemical potential [see Fig.~\ref{Fig_Gij_FSN_mu}(a,b)].  For this case, we have also calculated $G_{L(\pm)R(N)}(x_L)$ as a function of the position of the left lead while the position of the right one is fixed. We plot it together with quasiparticle spin and charge densities for the system in the  topological as well as nontopological phase, see Fig.~\ref{Sx_Q_GLR_3}. We note again that the sign of the nonlocal conductance is correlated with the sign of the local spin density, which holds across a wide range of lead attachment positions. We also note that when the  leads are attached to the ends of the system, the sign of the nonlocal conductance does not change  when the system undergoes a TPT since the local spin density has the same sign at the ends of the nanowire (see Fig.~\ref{Sx_Q_GLR_3} for details) in both topological and nontopological phases. Furthermore, we calculate nonlocal conductance maps $G_{L(\pm)R(N)}(\Delta_Z, E)$  (see Fig.~\ref{Fig_end}) in case where the leads are attached to the opposite ends of the nanowire. For such a setup one can observe that the  sign of the nonlocal conductance in the trivial ($\Delta_Z<\Delta_{sc}$) and in the topological ($\Delta_Z>\Delta_{sc}$) phase is the same. This highlights the importance of side coupling of the leads to the regions that are away from the nanowire ends.  Here, on purpose, we choose shorter nanowires  $L=3\ \mu$m  to illustrate clearly the case when the sign of the quasiparticle spin density $S_x(x)$ is position dependent [see orange curve  on~Fig.~\ref{Sx_Q_GLR_3}(b)]. The observed spin density oscillations and spin density sign change around the end of the nanowire, which occurs only in the topological phase, is determined essentially by the spin-orbit strength $\alpha$, Zeeman energy $\Delta_Z$ (more precisely by the induced gap $2|\Delta_{sc}-\Delta_{Z}|$), and the nanowire length $L$ which all affect the amplitude and frequency of oscillations and corresponding lengthscales (see App.~D for details). As a consequence, in this case the sign of the nonlocal conductances $G_{L(\pm)R(N)}(x_L)$ depends on the position of the left lead [see solid and dashed black curves on~Fig.~\ref{Sx_Q_GLR_3}(b)]. We expect that such a position dependent nonlocal conductance measurement of the quasiparticle spin (charge) density could be more easily achieved experimentally in 2D systems hosting spin polarized YSR states or vortex states using a pair of spin polarized (normal) STM tips.
 \begin{figure}[!t]
	\centering
	\includegraphics[width=8.6cm]{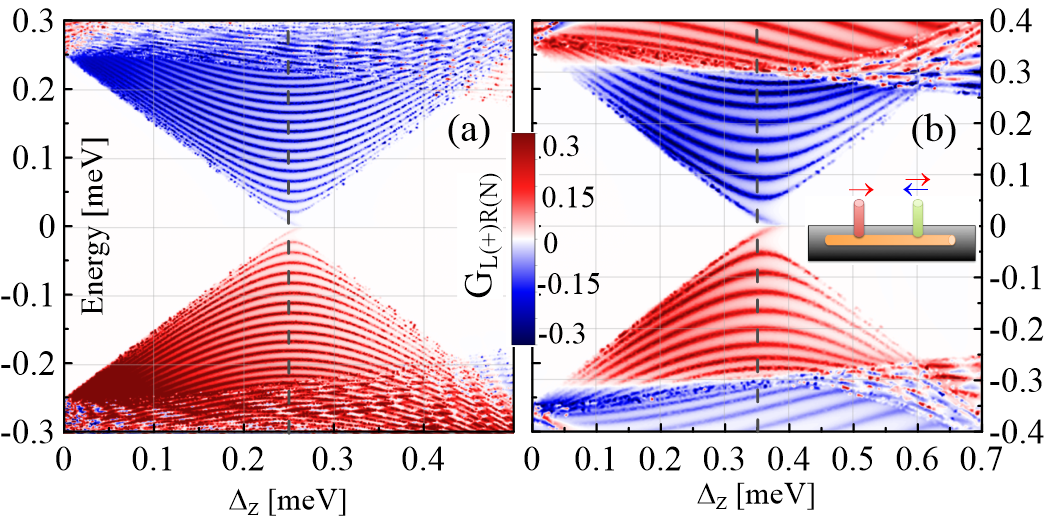}
	\caption{Maps of nonlocal conductance $G_{L(+)R(N)}(\Delta_Z, E)$ in the case when the leads are attached at  the opposite ends of the nanowire. Results shown in panel (a) are obtained for the same parameters as in Fig.~\ref{Fig_Gij_pFSN} with $L=10$~$\mu$m and $t_\Gamma=0.5t\approx0.14$~meV, while panel (b) illustrates results obtained for parameters from Fig.~\ref{Sx_Q_GLR_3} with $L=3$~$\mu$m $t_\Gamma=0.5t_c\approx1.5$ meV. Importantly, one can see clearly that the sign of the nonlocal conductance is the same for $\Delta_Z<\Delta_{sc}$ and for  $\Delta_Z>\Delta_{sc}$. Thus, it is independent of the spin density of the inner region of the system and consequently of the topological phase of the system. }
	\label{Fig_end}
\end{figure}
\onecolumngrid
\begin{center}
	\begin{figure}[!hb] 
		\centering
		\includegraphics[width=18cm]{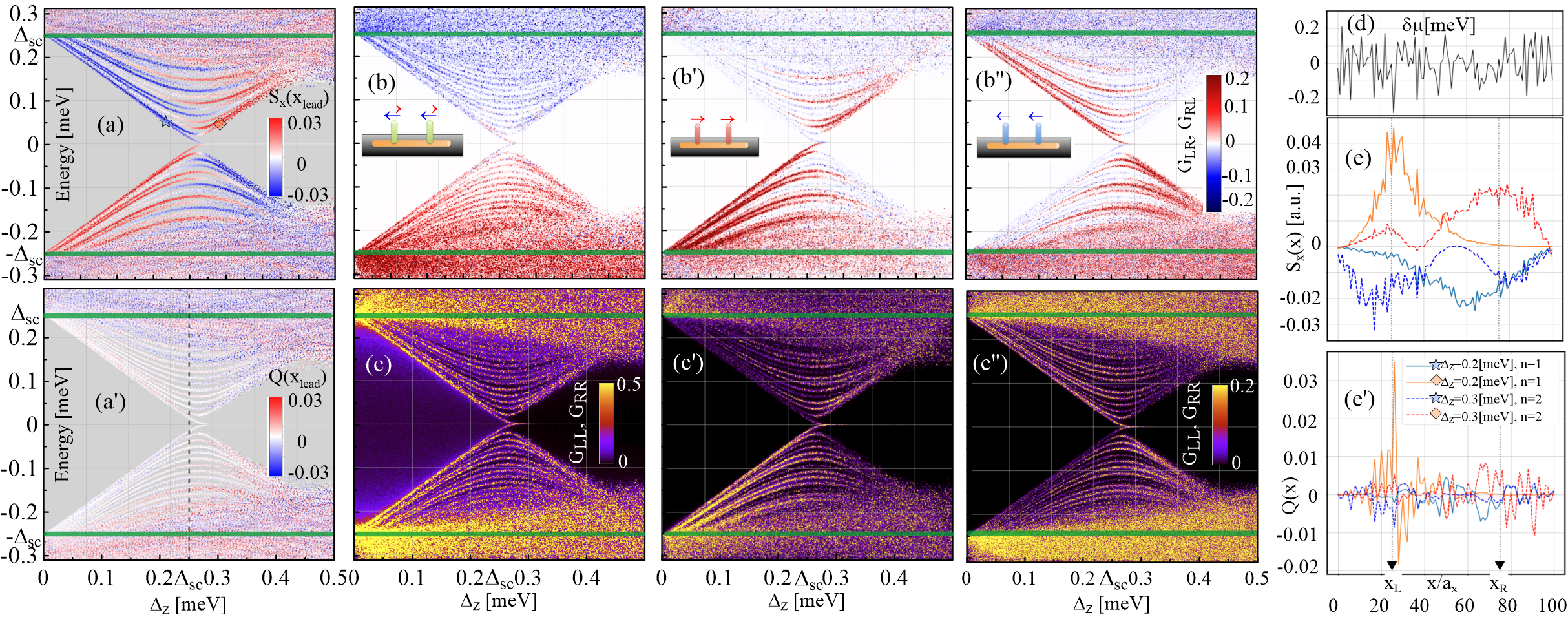}
		\caption{The same as in Fig.~\ref{Fig_Gii_Gij_NSN_FSF_Dz} but for system with static disorder $|\delta\mu_i|\le\Delta_{sc}$, the profile of which is depicted on panel (d).  Comparing to the clean system, the spin-dependent conductances are slightly affected, however, the main spin signature is still visible. (e) Even though the spatial profile of spin densities is significantly affected, the overall sign is preserved. (e') As expected, the biggest change can be seen in the charge density that takes much higher values comparable to the spin polarization, however, they are strongly oscillating around zero, which does not affect significantly the conductances.}
		\label{FSF_mu_dis}
	\end{figure}
\end{center}
\twocolumngrid
Finally, we show that our results are robust against disorder, by adding random on-site fluctuations to the onsite chemical potential $\mu_i=\mu+\delta\mu_i$ in Eq.~(1) with $|\delta\mu_i|\le\Delta_{sc}$ as shown in Fig.~\ref{FSF_mu_dis}(d) with corresponding mean free path $l_{mfp}~\simeq~685$~nm (see App. C for details). One can see that the presence of  disorder affects the energy levels [see Fig.~\ref{FSF_mu_dis}(a)] compared to the clean system [see Fig.~\ref{Fig_Gii_Gij_NSN_FSF_Dz}(a)], however, the sign of the local spin density is very robust to such disorder [see Fig~\ref{FSF_mu_dis}~(a,e)]. Moreover, as a consequence,  our transport study shows that the sign of the nonlocal conductance is also not affected by the disorder [see Fig.~\ref{Fig_Gij_pFSmF_Dz_nodis_dis}(a, b'), Fig.~\ref{Fig_Gij_pFSmF_mu_nodis_dis}(a,b'), Fig.~\ref{Fig_Gij_pFSmF_Dz_nodis_dis}(a,b'), Fig.~\ref{FSF_mu_dis}(b,b',b'',c,c',c'')]. Importantly,  in the charge neutrality regime where $\mu=0$ and where the band inversion is driven by the Zeeman energy, the spin-dependent nonlocal conductances are clearly more robust to the onsite disorder [see Fig.~\ref{Fig_Gij_pFSmF_Dz_nodis_dis}(a',b'), Fig.~\ref{FSF_mu_dis}(b,b', b'',c,c',c'')] than in the case when $\mu\ne0$ [see Fig.~\ref{Fig_Gij_pFSmF_mu_nodis_dis}(a',b')].

\onecolumngrid
\begin{center}
	\begin{table}[!ht]
	\centering
	\begin{tabular}{c||c||c||c||c||c|c||c|c||c|c||c|c}
		$S_x^{RNW}$  & \multicolumn{1}{c||}{$S_x<0${\color{blue} }}
		& \multicolumn{1}{|c||}{$S_x>0 $ }
		& \multicolumn{1}{|c||}{$S_x<0$ }
		& \multicolumn{1}{|c||}{$S_x>0$ }
		& \multicolumn{2}{|c||}{$S_x<0$}
		& \multicolumn{2}{|c||}{$S_x>0$ }		
		& \multicolumn{2}{|c||}{$S_x<0$ }
		& \multicolumn{2}{|c}{$S_x>0$ }\\ 
		& \multicolumn{1}{c||}{{\color{blue} $\leftarrow$}}
		& \multicolumn{1}{|c||}{${\color{red} \rightarrow}$ }
		& \multicolumn{1}{|c||}{ {\color{blue} $\leftarrow$}}
		& \multicolumn{1}{|c||}{${\color{red} \rightarrow}$ }
		& \multicolumn{2}{|c||}{ {\color{blue} $\leftarrow$}}
		& \multicolumn{2}{|c||}{${\color{red} \rightarrow}$ }		
		& \multicolumn{2}{|c||}{ {\color{blue} $\leftarrow$}}
		& \multicolumn{2}{|c}{${\color{red} \rightarrow}$ }\\ \hline
		Lead: & \multicolumn{1}{c||}{$L,R$:} & \multicolumn{1}{|c||}{$L,R$:} & 
		\multicolumn{1}{c||}{$L,R$:} & \multicolumn{1}{|c||}{$L,R$:} &
		\multicolumn{1}{c|}{$L$:} & \multicolumn{1}{c||}{$R$:} &
		\multicolumn{1}{c|}{$L$:} & \multicolumn{1}{c||}{$R$:}&
		\multicolumn{1}{c|}{$L$:} & \multicolumn{1}{c||}{$R$:} &
		\multicolumn{1}{c|}{$L$:} & \multicolumn{1}{c}{$R$:}		\\
		$G_{ij}$ / $S_{x}^{lead}$ &$S_x>0$ & $S_x>0$& 
		$S_x<0$ & $S_x<0$& 
		$S_x>0$ & $S_x<0$&
		$S_x>0$ & $S_x<0$& 
		$S_x>0$ & $S_x=0$&
		$S_x>0$ & $S_x=0$\\
		& {\color{red} $\rightarrow$} & {\color{red} $\rightarrow$} & 
		{\color{blue} $\leftarrow$} & {\color{blue} $\leftarrow$} & 
		{\color{red} $\rightarrow$} & {\color{blue} $\leftarrow$} &
		{\color{red} $\rightarrow$} & {\color{blue} $\leftarrow$} &
		{\color{red} $\rightarrow$} & {\color{teal} $\rightleftarrows$} &
		{\color{red} $\rightarrow$} & {\color{teal} $\rightleftarrows$} \\ \hline\hline
		$G_{LL}$   & \multicolumn{1}{c||}{Low}& \multicolumn{1}{|c||}{High}
		& \multicolumn{1}{|c||}{High}& \multicolumn{1}{|c||}{Low}
		& \multicolumn{2}{|c||}{Low}& \multicolumn{2}{|c||}{High}
		& \multicolumn{2}{|c||}{Low}& \multicolumn{2}{|c}{High}\\ \hline
		$G_{RR}$   & \multicolumn{1}{c||}{Low}& \multicolumn{1}{|c||}{High}
		& \multicolumn{1}{|c||}{High}& \multicolumn{1}{|c||}{Low}
		& \multicolumn{2}{|c||}{High}& \multicolumn{2}{|c||}{Low}
		& \multicolumn{2}{|c||}{$\approx 0$}& \multicolumn{2}{|c}{$\approx 0$}\\ \hline
		$G_{LR}$  & \multicolumn{1}{c||}{CAR/Low}& \multicolumn{1}{|c||}{ECT/High}
		& \multicolumn{1}{|c||}{ECT/High}& \multicolumn{1}{|c||}{CAR/Low}
		& \multicolumn{2}{|c||}{CAR/High}& \multicolumn{2}{|c}{ECT/Low}
		& \multicolumn{2}{|c||}{ECT/High}& \multicolumn{2}{|c}{CAR/High}\\ \hline
		$G_{RL}$   & \multicolumn{1}{c||}{CAR/Low}& \multicolumn{1}{|c||}{ ECT/High}
		& \multicolumn{1}{|c||}{ECT/High}& \multicolumn{1}{|c||}{CAR/High}
		& \multicolumn{2}{|c||}{ECT/Low}& \multicolumn{2}{|c}{CAR/High}
		& \multicolumn{2}{|c||}{CAR/Low}& \multicolumn{2}{|c}{CAR/High}\\ \hline
	\end{tabular}
	\indent
	\caption{Summary of results obtained for different  configurations of the spin polarization of the lowest nonzero energy state in a Rashba nanowire, $S_x^{RNW}$, and the spin polarization  of the leads, $S_x^{L,R}$, and their correspondence with the value of the local ($G_{ii}$) and nonlocal conductance ($G_{ij}$), $i,j=L,R$. The table also contains  information about the leading processes contributing to the nonlocal conductance, being either ECT or CAR, corresponding to the positive or negative sign of the nonlocal conductance,  respectively. }
	\label{tab_1}
\end{table}
\end{center}
\twocolumngrid
\noindent
As a summary, in Table~\ref{tab_1}, we list schematically all the considered spin configurations of the left and right leads together with spin polarization of the lowest nonzero energy states in the Rashba nanowire and the corresponding information about the sign 
(distinguishing between CAR or ECT dominating channels) as well as the strength of the local and nonlocal conductance signals.

\subsection{Quasiparticle charge detection}

For the goal of detecting the quasiparticle charge, we consider a setup with two normal (unpolarized) leads. First, we start with calculating the local quasiparticle charge $Q(E_n)$ as a function of Zeeman energy $\Delta_Z$ for states at a given energy. The energy spectrum of the finite-size system is depicted in the Fig.~\ref{Fig_Gii_Gij_NSN_FSF_Dz}(b) where the blue/red color indicates the negative/positive sign of the quasiparticle charge $Q(E_n)$ for a given energy eigenstate at selected position at which the normal lead  will be attached in case of transport simulations. Here, we consider the charge neutrality regime where $\mu=0$. In such a case, both before ($\Delta_Z<\Delta_{sc}$) and after ($\Delta_Z>\Delta_{sc}$) the TPT, the quasiparticle charge is very close to zero. In addition,  if one looks closely at the two lowest energy quasiparticle, one finds that the corresponding quasiparticle charge density along the nanowire is also almost zero with some small oscillations at the nanowire ends as depicted in the Fig.~\ref{Fig_Gii_Gij_NSN_FSF_Dz}(e) and Fig.~\ref{Sx_Q_GLR_3}. This means that the low-energy quasiparticles are composed of approximately equal amount of  particle and hole parts. This is also consistent with the analytical predictions for $k\approx0$, $Q(k)\approx sign(\Delta_Z-\Delta_{sc})[\hbar^2k^2/(2m\Delta_{sc})]$ (see App.~A). In such a scenario, when the two leads attached to the system are normal (unpolarized), there is no visible change in sign of the nonlocal conductance $G_{L(N)R(N)}(\Delta_{Z}, E)=G_{R(N)L(N)}(\Delta_Z, E)$ [see Fig.~\ref{Fig_Gii_Gij_NSN_FSF_Dz}(b)] before and after the TPT.

On the other hand, the system can be tuned away from the charge neutrality point ($\mu=0$), e.g. when the TPT (the band inversion) is driven by the change of chemical potential $\mu$ while the Zeeman energy is fixed to  a value that is greater than the induced superconducting gap, i.e. $\Delta_Z>\Delta_{sc}$. Here, in contrast to the previously considered regime, the quasiparticle charge is generally nonzero for a wide range of parameters and, more importantly, it changes its sign [see Fig.~\ref{Fig_mu}(a')] around the critical values of the chemical potential $\mu_c^\pm=\pm\sqrt{\Delta_Z^2-\Delta_{sc}^2}$ for which the energy gap in the spectrum closes and the Andreev bands get inverted. As a consequence, the  sign of the nonlocal conductance $G_{L(N)R(N)}(\mu, E)$ is flipped accordingly around the TPT points $\mu_c^\pm$ ~[see Fig.~\ref{Fig_mu}(b)]. This is in general agreement with theoretical~\cite{Nonlocal_BCS_charge_th,Tunable_car_PRL_theory} and experimental studies~\cite{Tunable_car_PRX}.

\section{Conclusions}

We analyzed in detail spin and transport properties of a proximitized Rashba nanowire in a three-terminal setup with  grounded superconductor and with tunnel-coupled side normal or spin polarized leads. We have considered several configurations of the spin polarization of the leads, revealing a distinct correspondence between the sign of the nonlocal conductance and the sign of the local quasiparticle spin density. In particular, in the setups featuring two leads with the same or opposite spin polarizations, we observed a dominance of either the ECT
or CAR
processes, if the spin of the probed quasiparticle state matches the spin of the injected charge carrier in the lead. Alternatively, employing one normal and one spin-polarized lead facilitated the precise mapping between the sign of the nonlocal conductance and the spin polarization of the probed quasiparticle state. Furthermore, nonlocal conductances in setups with two normal leads provide information solely pertaining to the quasiparticle charge, rather than spin. We showed that such a behavior can be used to detect the TPT which involves  band inversion and a related sign inversion of spin and charge of the lowest energy states.
Moreover, we showed that a Rashba nanowire with tunnel-coupled side leads can be a versatile platform for tuning between CAR and ECT processes. This functionality holds promise for applications such as  Cooper pair splitters and minimal Kitaev chain systems of quantum dots hosted in proximitized Rashba nanowires.
Importantly, our findings highlight the importance of coupling leads to regions away from the nanowire ends, which gives additional insight into the charge and spin density characteristics of quasiparticles.  We expect that our results will be particularly useful to experimentalists working on hybrid superconductor-semiconductor systems.

While our study focused on spin-polarized leads, we expect analogous outcomes for systems featuring spin-polarized quantum dots~\cite{spin_filter_delftNC, spin_filter_dots_Marcus}. However, this case can  add complexity in tuning dot levels to match the energy levels of the probed quasiparticle states and taking into account Coulomb charging physics in the quantum dots. In summary, our work presents a novel avenue for detecting quasiparticle spin polarization and studying CAR and ECT dominated regimes through measuring the spin density profiles of the lowest energy Andreev bound states with tunnel-coupled side normal or spin polarized leads.

\section{ACKNOWLEDGMENTS} 
We acknowledge helpful discussions with Andreas Baumgartner. This work was supported within POIR.04.04.00-00-5CE6/18 project carried out within the HOMING programme of the Foundation for Polish Science co-financed by the European Union under the European Regional Development Fund and Swiss National Science Foundation. We gratefully acknowledge the Polish high-performance computing infrastructure PLGrid (HPC Centers: ACK Cyfronet AGH) for providing computer facilities and support within the computational grant no. PLG/2023/016184. This work was supported by  the Swiss National Science Foundation. This project received funding from the European Union's Horizon 2020 research and innovation program (ERC Starting Grant, Grant No 757725). 
\appendix
\section{Hamiltonian in momentum space}
The bulk energy bands  of the system can be studied in momentum space by imposing periodic boundary conditions. In order to write ${H}$ in the momentum space one can use Fourier transformed operators ${c}_{j \sigma}=\sum_j {c}_{k \sigma} e^{-ijka}/\sqrt{N}$, and ${\Psi}_k~=~({c}_{k \uparrow}, {c}_{k \downarrow}, {c}_{-k \downarrow}^{\dag}, -{c}_{-k \uparrow}^{\dag})^T$. The corresponding Hamiltonian takes the form ${H}~=~\sum_k{\Psi}_k^\dag{\mathcal H}(k){\Psi}_k$ with

\begin{align}
&  {\mathcal H}(k)=\left[2t-2t\cos(ka)-\mu+2\tilde{\alpha}\sin(ka)\sigma_y\right]	\tau_z \nonumber\\
&\hspace{100pt}+\Delta_{sc}\tau_x+\Delta_Z\sigma_x. \label{hamiltonian_momentum_tb}
\end{align} 
and in the continuum limit $ka\ll 1$ \cite{composite_majorana},
we get 
\begin{equation}\label{hamiltonian_momentum_open}
{\mathcal H}(k)=\left(\frac{\hbar^2k^2}{2m} -\mu+\alpha k \sigma_y\right)\tau_z+\Delta_{sc}\tau_x+\Delta_{Z}\sigma_x.
\end{equation}

The continuum model and its discretized tight binding version are related by the hopping amplitude $t=\hbar^2/(2ma^2)$~\cite{Diego_2013}. Regarding the parameters, we use the same notation as in the main text. By diagonalizing ${\mathcal H}(k)$ [see Eq.  ($\ref{hamiltonian_momentum_tb}$) or ($\ref{hamiltonian_momentum_open}$)], we obtain analytical expressions for the eigenvalues  $E_{RNW}^{\lambda\eta}(k)$ and corresponding eigenstates $\Phi_{\lambda\eta}(k)$ whose explicit forms can be found in the supplemental material to Ref.~\cite{Spin_and_charge}. As a results one gets four energy bands (see $E_{RNW}(k)$ in the middle panels of Fig.~\ref{bands}), labeled by  $\lambda$ and $\eta$, where $\lambda=1$ ($\lambda=\bar 1$) labels bands with positive (negative) energy and $\eta=\bar 1$ the bands closest to the Fermi level. However, in the main text for the purpose of clarity we skipped energy band labels $\lambda\eta$. The corresponding bulk quasiparticle spin and charge can be calculated as follows:
\begin{align}
&{S}_x^{\lambda\eta}(k)~=~\Phi_{\lambda\eta}^\dag(k){\sigma}_x\Phi_{\lambda\eta}(k),\label{spin_formula_k}\\
&Q_{\lambda\eta}(k)~=~-\Phi_{\lambda\eta}^\dag(k)\tau_z\sigma_0\Phi_{\lambda\eta}(k),\label{charge_formula_k}
\end{align}
which for $k\approx0$ take the approximate analytical forms~\cite{Spin_and_charge}
\begin{align}
&S_x^{\lambda\bar 1}(k)\approx\lambda\textrm{sign}(\Delta_Z-\Delta_{sc})\Big[1-\frac{(\alpha k)^2}{2(\Delta_Z-\Delta_{sc})^2}\Big] \label{spin_analytic}\, ,\\
&Q_{\lambda\bar 1}(k)\approx\lambda\textrm{sign}(\Delta_Z-\Delta_{sc})\frac{\hbar^2k^2}{2m\Delta_{sc}}\label{charge_analytic}\, .\\\notag
\end{align}

\noindent
\section{Shorter Wire Limit}
\begin{figure}[!ht] 
	\centering
	\includegraphics[width=8.6cm]{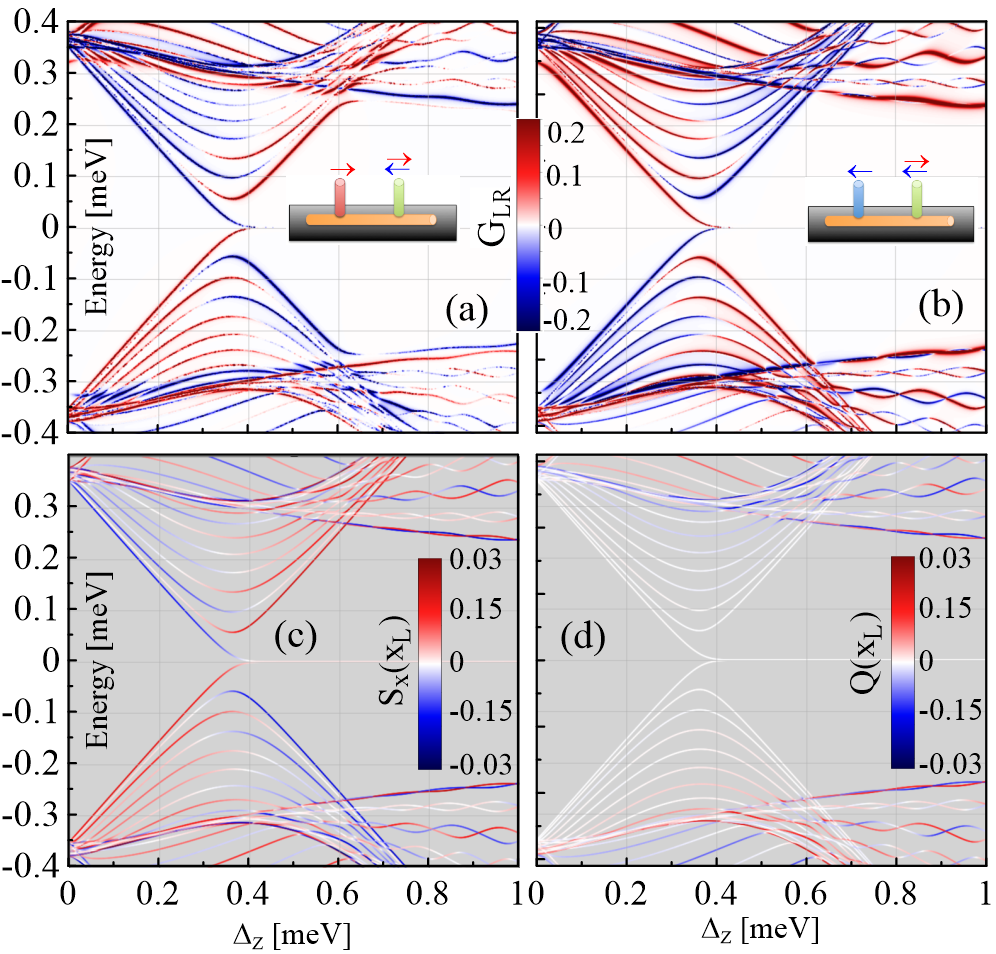}
	\caption{
		The	maps of nonlocal conductance (a) $G_{L(+)R(N)}(\Delta_Z, E)$, (b) $G_{L(-)R(N)}(\Delta_Z, E)$ for the nanowire of length $L=3$~$\mu$m ($N=100$, $a=30$~nm, $t\approx3$ meV, $t_\Gamma=0.15t\approx0.45$ meV). The energy spectrum of the nanowire together with (c) local quasiparticle spin $S_x(E, x_L)$  and (d) local quasiparticle charge $Q(E, x_L)$  at $x_L=45a$. Here, the superconducting gap in the nanowire is set to $\Delta_{sc}=0.35$~meV and $\alpha=40$ meVnm ($E_{\rm SO}=0.15$ meV). Again, there is  perfect correspondence between  the sign of local spin of the probed quasiparticles [see panel (c)] and the sign of  nonlocal conductance $G_{LR}(\Delta_Z, E)$ [see panels (a,b)].}
	\label{Fig_Gij_FSN_Dz_a30}
\end{figure}
\begin{figure}[!ht] 
	\centering
	\includegraphics[width=8.6cm]{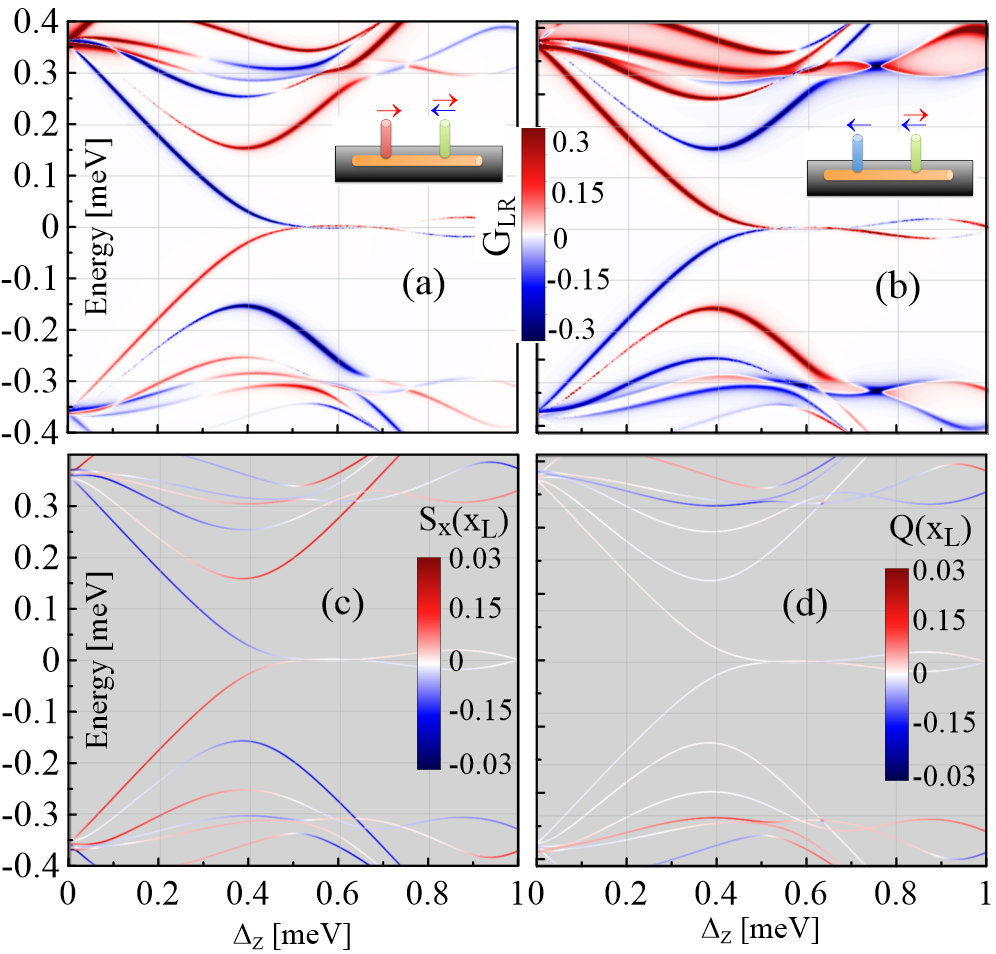}
	\caption{The same as Fig.~\ref{Fig_Gij_FSN_Dz_a30} but for the length $L=1$~$\mu$m ($N=100$, $a=10$ nm, $t\approx27.21$ meV, $t_\Gamma=0.15t\approx4.1$ meV). In a short nanowire, in the topological phase, we also get an oscillating signal around zero energy from two overlapping Majorana bound states.}
	\label{Fig_Gij_FSN_Dz_a10}
\end{figure}

In this Appendix,  we consider the setup with left lead being spin up/down polarized and the right lead being normal. The corresponding nonlocal conductances $G_{L(\pm)R(N)}(\Delta_Z, E)$ for the nanowire of the length $L=3$~$\mu$m ($N=100$, $a=30$~nm, $t\approx3$~meV, $t_\Gamma=0.15t\approx0.45$ meV, $x_L/a=45$, $x_R/a=55$) and $L=1$~$\mu$m ($N=100$, $a=10$~nm, $t\approx27.21$ meV, $t_\Gamma=0.15t\approx4.1$~meV, $x_L/a=45$, $x_R/a=55$) are presented in Fig.~\ref{Fig_Gij_FSN_Dz_a30}(a,b) and 
Fig.~\ref{Fig_Gij_FSN_Dz_a10}(a,b), respectively. Here, we choose  the distances between the gates as $d=x_R-x_L=300$~nm and $d=x_R-x_L=100$~nm. The corresponding energy spectrum calculated as a function of Zeeman energy $\Delta_Z$  for the proximitized nanowire is shown in Fig.~\ref{Fig_Gij_FSN_Dz_a30}~(c,d) and

 Fig.~\ref{Fig_Gij_FSN_Dz_a10}~(c,d). On panels labeled by (c)  [(d)]  of Fig.~\ref{Fig_Gij_FSN_Dz_a30} and Fig.~\ref{Fig_Gij_FSN_Dz_a10}, the color represents the sign and strength of quasiparticle spin [charge]. There is again a perfect correspondence between the sign of the nonlocal conductance [panels (a,b)] and the sign of the local quasiparticle spin [panel (c)]. Here, however, in the topological phase, one notices the oscillations around zero energy resulting from the overlap of two Majorana bound states localized at the opposite ends of the nanowire. These oscillations manifests themselve also in nonzero nonlocal conductance signal in short Rashba nanowires in which the Majorana bound state has non-zero support at points at which we attach the leads.
\begin{figure}[!hb] 
	\centering
	\includegraphics[width=8.1cm]{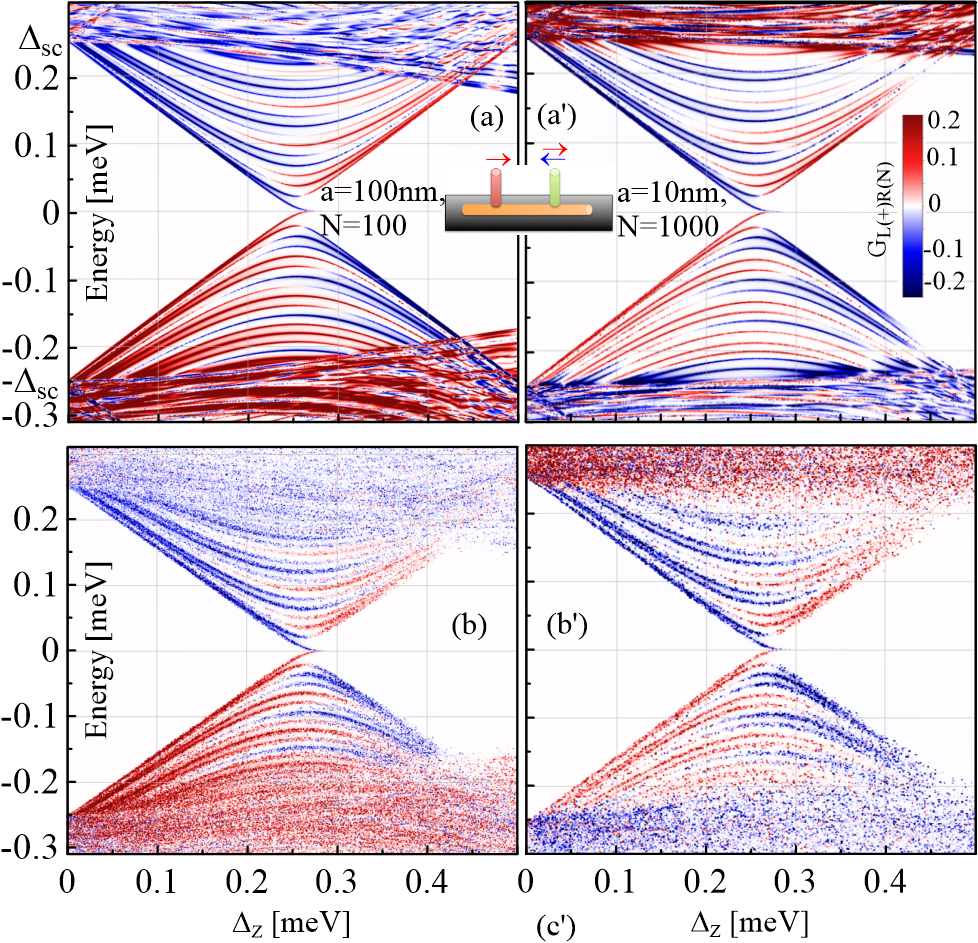}
	\caption{The comparison of the nonlocal conductance $G_{L(+)R(N)}(\Delta_Z, E)$ obtained for different lattice parameters for the same nanowire length $L=10$~$\mu$m. Results for $a=100$~nm and $N=100$  [as in the main text, see Fig.~\ref{Fig_Gij_pFSN}(a)] and for $a=10$ nm and $N=1000$ are presented on panel (a) and (a'), respectively. In the regime of interest, $|E|<\Delta_{sc}$, the obtained results are almost identical. Panels (b) and (b') illustrate nonlocal conductances for systems with  disorder, which is depicted correspondingly in Fig.~\ref{Fig_Gii_Gij_NSN_FSF_Dz}(d) and Fig.~\ref{Fig_Gii_Gij_NSN_FSF_Dz_a10}(c). }
	\label{Fig_pFSN_comp}
\end{figure}
\section{Lattice spacing, numerical check}
\begin{center}
	\begin{figure}[]
		\centering
		\includegraphics[width=17.5cm]{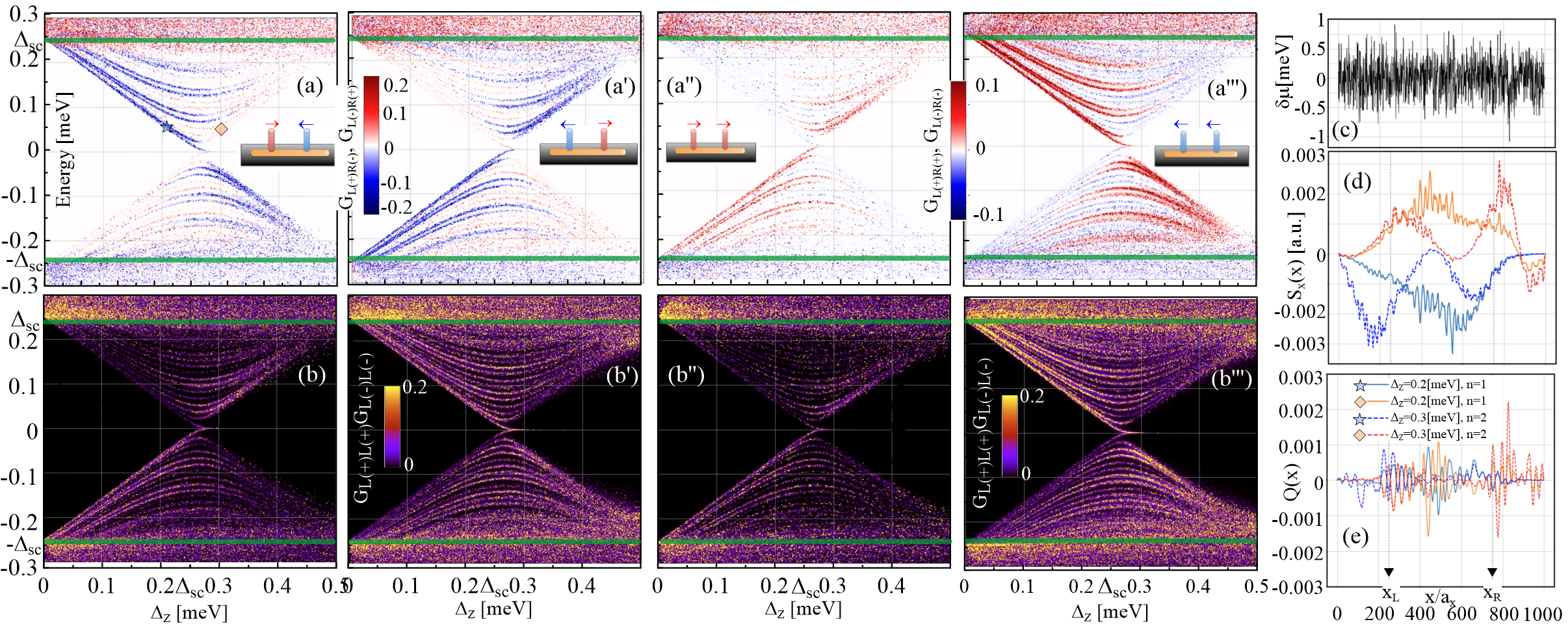}
		\centering
		\caption{The nonlocal conductance maps for various configurations of spin polarization in the leads are shown in the following panels: (a) $G_{L(+)R(-)}(\Delta_Z, E)$, (a') $G_{L(-)R(+)}(\Delta_Z, E)$, (a'') $G_{L(+)R(+)}(\Delta_Z, E)$, (a''') $G_{L(-)R(-)}(\Delta_Z, E)$. The corresponding local conductance maps $G_{L(\pm)L(\pm)}(\Delta_Z, E)$ are shown on the lower panels (b-b'''). The onsite disorder is presented in panel (c), while the spin and charge densities for selected states, marked by blue and red symbols, are depicted in panels (d) and (e), respectively, similar to  Fig.~\ref{FSF_mu_dis} in the main text.}
		\label{Fig_Gii_Gij_NSN_FSF_Dz_a10}
	\end{figure}
\end{center}

 As we mentioned in the main text, for the purpose of numerical efficiency, we set the effective tight binding lattice spacing to $a=100$ nm in order to study the long wire limit ($L=10$ $\mu$m with $N=100$). Here, we show that, for $10$ times smaller lattice spacing $a=10$ nm ($L=10$~$\mu$m with $N=1000$, $t\approx27.21$ meV, $t_\Gamma=0.2t\approx5.45$ meV), the key results are almost identical. As an example, we consider the setup with  left lead being spin-up polarized and right  lead being unpolarized and compare directly the corresponding nonlocal conductance map $G_{L(+)R_(N)}(\Delta_Z, E)$ for the parameters $a=100$~nm ($N=100$) and $a=10$~nm  ($N=1000$), see Fig.~\ref{Fig_pFSN_comp}. 
Furthermore, we investigate whether the presence of disorder affects the results similarly when using two different lattice constants: $a=100$ nm ($N=100$) and $a=10$ nm ($N=1000$) while keeping the nanowire length constant at $L=10$~$\mu$m. In the latter case we set even stronger disorder $|\delta\mu_i|\lesssim 1$ meV (Gaussian distribution with $\sigma=0.3$ meV) to have similar mean free paths in both cases (see appropriate comment at the end of this Appendix). It can be observed in Fig.~\ref{Fig_pFSN_comp} (b,b') that the nonlocal conductance signature of the Andreev band spin remains clearly visible also for the smaller lattice constant. For completeness,  we also calculate maps of local $G_{LL}(\Delta_Z, E)$ and nonlocal conductance $G_{LR}(\Delta_Z, E)$ for other configurations of spin polarized leads: L(+)R(-), L(-)R(+), L(+)R(+), L(-)L(-) which are illustrated in Fig.~\ref{Fig_Gii_Gij_NSN_FSF_Dz_a10} for lattice spacing $a=10$ nm ($L=10$~$\mu$m with $N=1000$, $t\approx27.21$ meV, $t_\Gamma=0.2t\approx5.45$ meV) and disorder amplitude $|\delta\mu_i|\lesssim 1$ meV. In order to estimate the mean free path $l_{mfp}$ induced by the onsite electrostatic disorder $\delta\mu_i$ one can use Fermi's golden rule which gives~\cite{Thamm_2024, Cole_2016, Awoga_2023}:
 	\begin{align}
 	l_{mfp}=v_F\tau \simeq \frac{3\hbar v_F}{2\pi a N_{1D}(0) W^2}
 	\end{align}
 	where $N_{1D}(0)=2/(\pi\hbar v_F)$ is the 1D density of states, and W the disorder strength corresponding to the amplitude of fluctuations of the onsite chemical potential  $-W<\delta\mu_i<W$. This yields the following estimates for the mean free path: $l_{mfp}\simeq 685$~nm ($l_{mfp}\simeq 617$~nm) for $a=100$~nm and $W=0.3$~meV ($a=10$~nm and $W=1$~meV).
\onecolumngrid
\twocolumngrid
\noindent
 \begin{figure}[!ht]
	\centering
	\includegraphics[width=8cm]{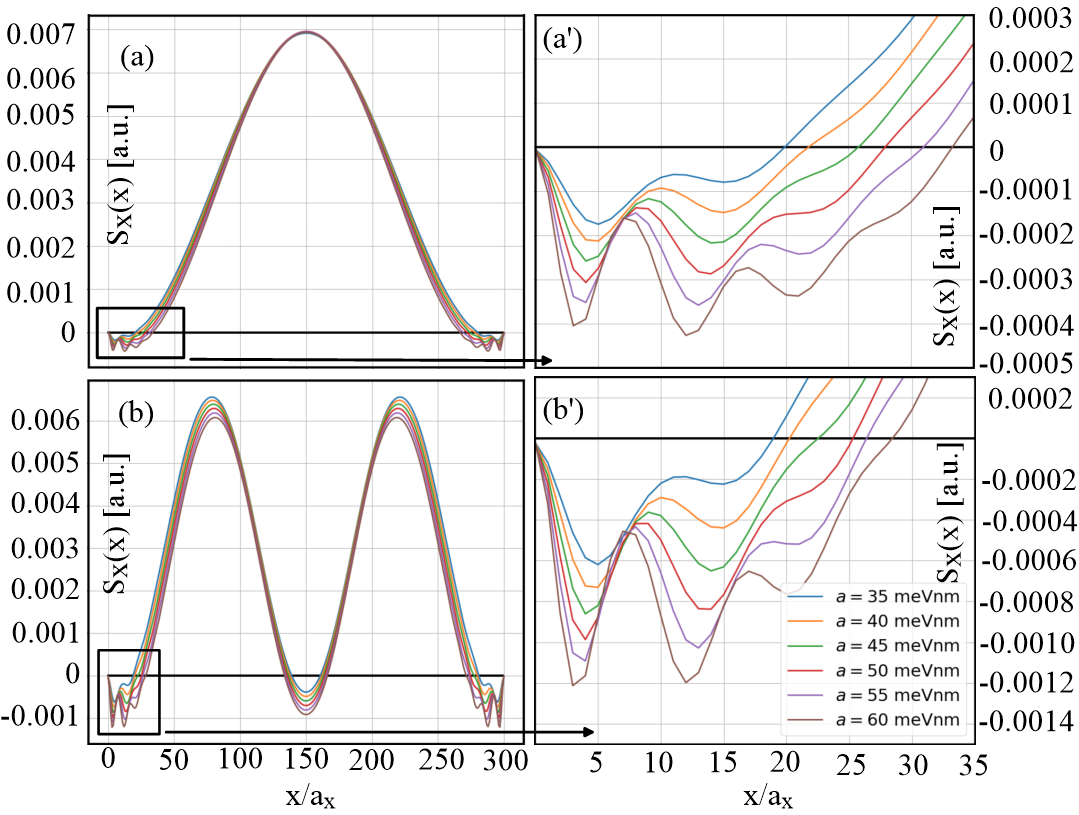} 
	\caption{The $x$ component of spin the density $S_x(x)$ as a function of position along the nanowire for the lowest and second positive energy states are shown in panels (a) and (b), respectively. The primed panels (a'), (b') show a zoom-in of the $S_x$ oscillations at the left end of the nanowire. Colors denote different values of the spin-orbit strength $\alpha$ (as indicated by the legend), which affects the amplitude and frequency of oscillations, as well as the position where the spin density changes sign from positive to negative. Here, we set $L=9$ $\mu$m, $a=30$ nm, $N=300$, $\Delta_Z=0.4$ meV, $\Delta_{sc}=0.35$ meV.}
	\label{Fig_alpha1}
\end{figure}

\section{Spin density profile at nanowire ends}
 \begin{figure}[!hb]
	\centering
	\includegraphics[width=8.6cm]{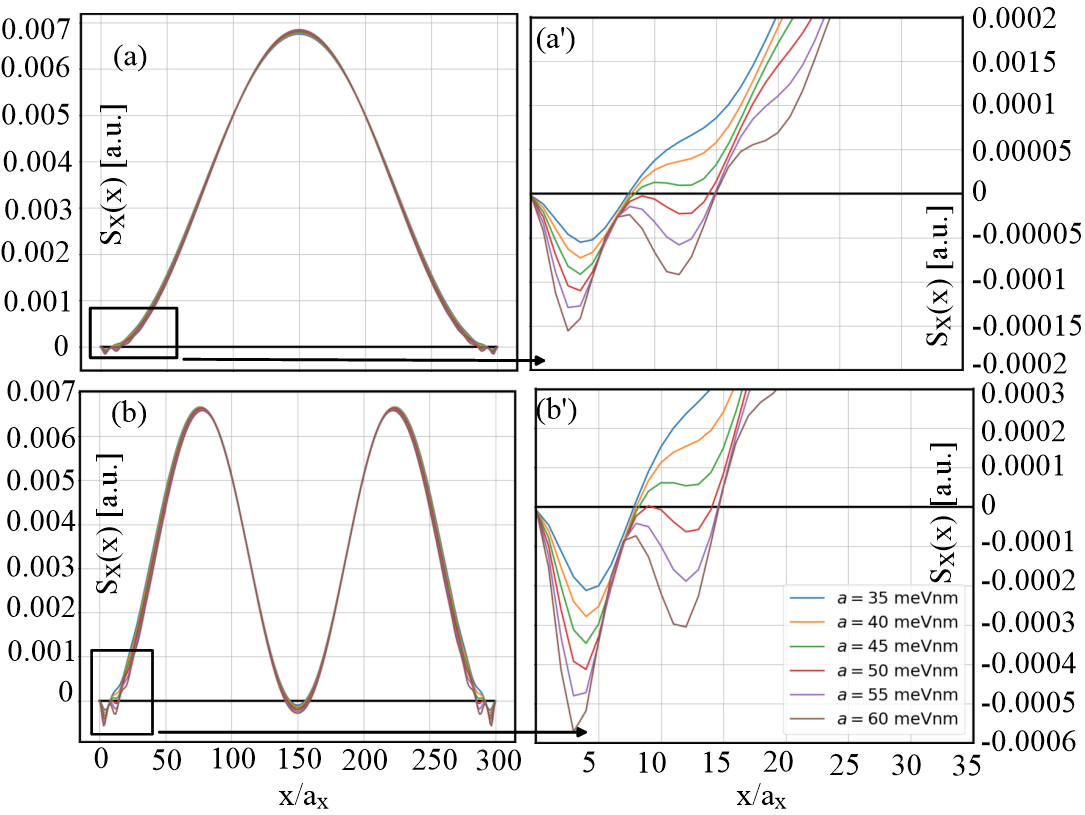}
	\caption{Same as Fig.~\ref{Fig_alpha1} but for larger Zeeman energy set to $\Delta_Z=0.45$ meV which results in a larger induced gap ($\Delta_i=2|\Delta_{sc}-\Delta_Z|=0.2$ meV). This larger induced gap suppresses the amplitude of spin density $S_x(x)$ oscillations at the nanowire ends.}
	\label{Fig_alpha2}
\end{figure}
\begin{figure}[!ht]
	\centering
	\includegraphics[width=8.6cm]{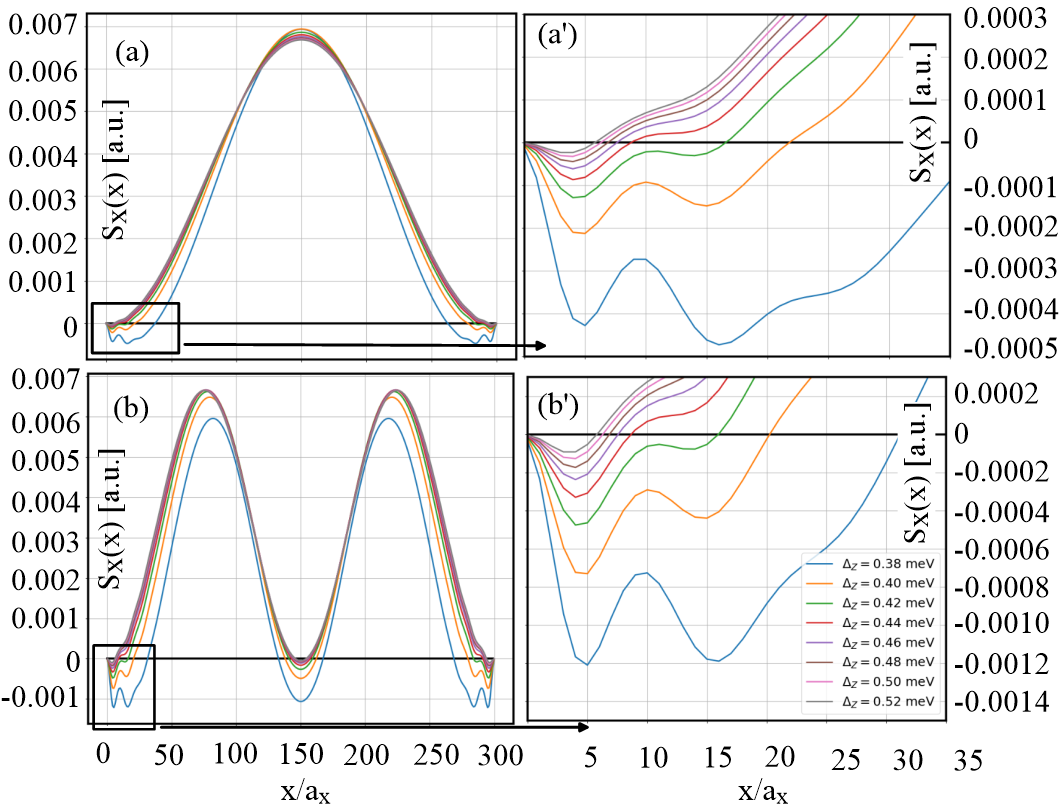}
	\caption{Similar to Figs.~\ref{Fig_alpha1}~and~\ref{Fig_alpha2} but for different values of Zeeman energy ranging from $\Delta_Z=0.38$~meV to $\Delta_Z=0.52$~meV. The spin-orbit strength is fixed at $\alpha=40$~meVnm. The colors corresponds to the spin densities $S_x(x)$ plotted for different values of Zeeman energy $\Delta_Z$ (as indicated by legend).}
	\label{Fig_Dz}
\end{figure}
\begin{figure}[!hb]
	\centering
	\includegraphics[width=8.6cm]{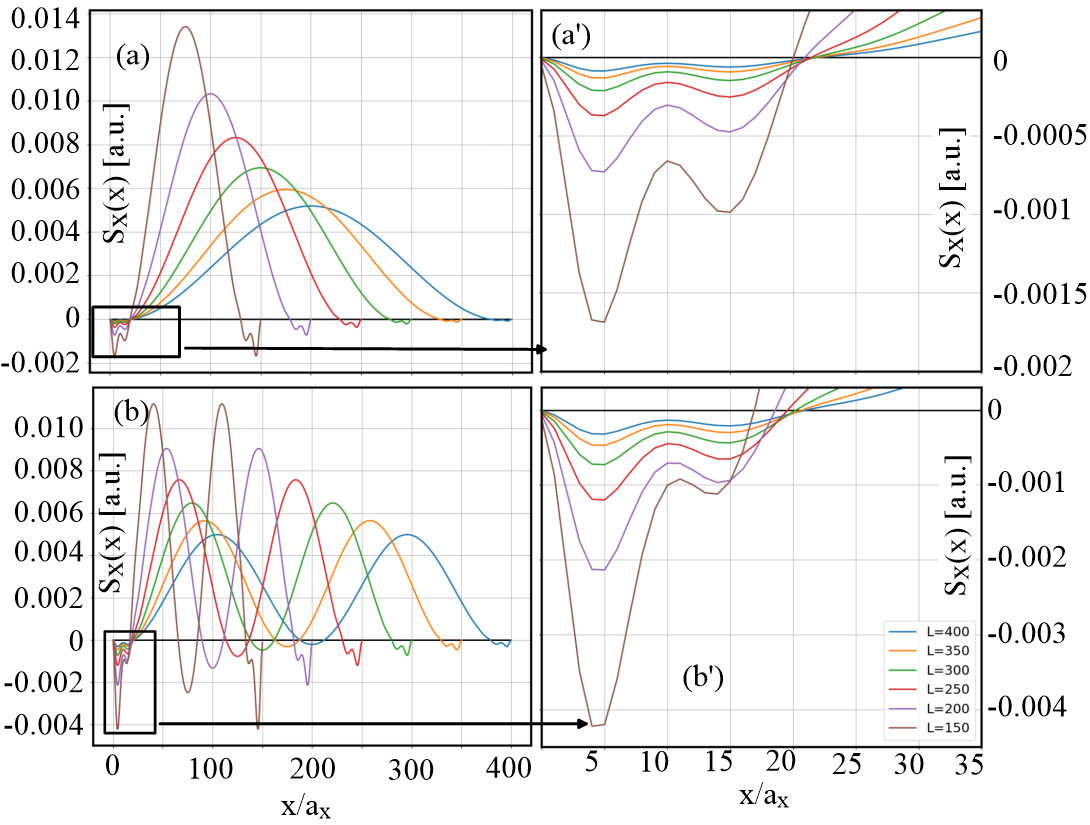}
	\caption{Similar to previous figures Figs. (\ref{Fig_alpha2}-\ref{Fig_Dz}) but for different lengths of the system ranging between $L=4.5$ $\mu$m to $L=12$ $\mu$m. The spin-orbit strength is fixed at $\alpha=40$ meVnm and the Zeeman energy to $\Delta_Z=0.45$ meV. The colors corresponds  to the spin densities $S_x(x)$ for these different system lengths. One can clearly see that the amplitude of spin density oscillations decreases as the length of the system increases. }
	\label{Fig_L}
\end{figure}

Here, we examine how the system parameters - Zeeman energy $\Delta_Z$, spin-orbit strength $\alpha$, and length of the system $L$ affect the spin density profile $S_x(x)$ at the ends of the nanowire for the two lowest positive energy bulk states. To be specific, we first set the parameters to $L=9$~$\mu$m, $a=30$ nm, $N=300$, $\Delta_{sc}=0.35$ meV, $\Delta_{Z}=0.4$ meV such that the  system is near the TPT and the induced gap is small. We then plot the spin densities $S_x(x/a_x)$ along the wire for the first two  energetically lowest  states (with positive energy) varying the spin-orbit strength from $\alpha=35$ meVnm to $\alpha=60$ meVnm (see Fig.~\ref{Fig_alpha1}). One can observe that both the amplitude and frequency of the spin density oscillations increase with $\alpha$.  Next, we increase the Zeeman energy to $\Delta_{Z}=0.45$ meV which moves the system further away from the TPT and makes the induced gap larger. As shown in Fig.~\ref{Fig_alpha2} the amplitude of $S_x(x)$ oscillations is smaller than in the previous case.  In the subsequent analysis, we fix the spin-orbit strength to $\alpha=40$~meVnm and vary the Zeeman energy between $\Delta_Z=0.38$ meV and $\Delta_Z=0.52$. The corresponding Fig.~\ref{Fig_Dz} demonstrates that the amplitude, frequency, and the position where the spin density changes sign are strongly dependent on $\Delta_Z$.  We expect that the oscillation and sign change of the spin density, which occur only in the topological phase, reflect the presence of a zero-energy Majorana state in the system's energy spectrum, whose wave function oscillates around zero at the nanowire end. Most likely, these oscillations arise from the interplay between interior and exterior energy gaps~\cite{composite_majorana}. Finally, we investigate how the length of the system affects spin density oscillations. Fig.~\ref{Fig_L} shows that the length of the system $L$  primarily influences the amplitude of the oscillations: the amplitude increases as the length decreases. This effect is likely related to the hybridization of Majorana states, which becomes more pronounced with a shorter wire. The observation that the spin density changes sign near the end only in the topological phase is probably due to the presence of Majorana states in the energy spectrum, which also affects the spin density of other energy states. This behavior could potentially serve as a new hallmark of the topological superconducting phase, however, a more detailed analysis is beyond the scope of this work and could be a topic for future research.

\end{document}